\documentclass[longauth]{aa}
\usepackage{graphicx}
\usepackage{amsmath}	
\usepackage{amssymb}	
\usepackage{bm}		
\usepackage{booktabs}
\usepackage{listings}
\usepackage{color}
\usepackage{threeparttable}
\usepackage{enumerate}
\usepackage{indentfirst} 
\usepackage{stackengine}
\newcommand\xrowht[2][0]{\addstackgap[.5\dimexpr#2\relax]{\vphantom{#1}}}
\usepackage{booktabs}
\usepackage{txfonts}
 \usepackage[bookmarks=false,         
     pdfnewwindow=true,      
     colorlinks=true,    
     linkcolor=blue,     
     citecolor=blue,     
     filecolor=blue,  
     urlcolor=blue,      
final=true
 ]{hyperref}
 \usepackage{etoolbox}
 \usepackage{color}
\usepackage{CJK}

\newcommand{\xmm}{{\emph{XMM-Newton}}}
\newcommand{\nustar}{{\emph{NuSTAR}}}
\newcommand{\nodata}{\centering $\cdot \cdot \cdot$}

\begin{document} 
\begin{CJK*}{UTF8}{gbsn}

   \title{Transient obscuration event captured in NGC~3227 \\
   II. Warm absorbers and obscuration events in archival \emph{XMM-Newton} and \emph{NuSTAR} observations}

   \author{Yijun~Wang (王倚君)
          \inst{1,2,3,4,5,6},
          Jelle~Kaastra
          \inst{2,1},
          Missagh~Mehdipour
          \inst{10,2},
          Junjie~Mao (毛俊捷)
          \inst{7,8,2},
          Elisa~Costantini
          \inst{2,9},
          Gerard~A.~Kriss
          \inst{10},
          Ciro~Pinto
          \inst{11},
          Gabriele~Ponti
          \inst{12,13},
          Ehud~Behar
          \inst{14},
          Stefano~Bianchi
          \inst{15},
          Graziella~Branduardi-Raymont
          \inst{16},         
          Barbara~De~Marco
          \inst{17},
          Sam~Grafton-Waters
          \inst{16},
          Pierre-Olivier~Petrucci
          \inst{18},
          Jacobo~Ebrero
          \inst{19},
          Dominic~James~Walton
          \inst{20},
          Shai~Kaspi
          \inst{21},
          Yongquan~Xue (薛永泉)
          \inst{3,4},
          St\'ephane~Paltani
          \inst{22},
          Laura~di~Gesu
          \inst{23},
          Zhicheng~He (何志成)
          \inst{3,4}
          }

   \institute{Leiden Observatory, Leiden University, Niels Bohrweg 2, 2300 RA Leiden, The Netherlands\\
              \email{Y.Wang@sron.nl}
         \and SRON Netherlands Institute for Space Research, Niels Bohrweg 4, 2333 CA Leiden, The Netherlands
         \and CAS Key Laboratory for Research in Galaxies and Cosmology, Department of Astronomy, University of Science and Technology of China, Hefei 230026, China
         \and School of Astronomy and Space Science, University of Science and Technology of China, Hefei 230026, China
         \and Department of Astronomy, Nanjing University, Nanjing 210093, China
         \and Key Laboratory of Modern Astronomy and Astrophysics (Nanjing University), Ministry of Education, Nanjing 210093, China
         \and Department of Physical, Hiroshima University, 1-3-1 Kagamiyama, HigashiHiroshima, Hiroshima 739-8526, Japan
         \and Department of Physics, University of Strathclyde, Glasgow G4 0NG, UK
         \and Anton Pannekoek Astronomical Institute, University of Amsterdam, P.O. Box 94249, 1090 GE Amsterdam, the Netherlands
         \and Space Telescope Science Institute, 3700 San Martin Drive, Baltimore, MD 21218, USA
         \and INAF-IASF Palermo, Via U. La Malfa 153, I-90146 Palermo, Italy
         \and INAF-Osservatorio Astronomico di Brera, Via E. Bianchi 46, 23807 Merate, LC, Italy
         \and MAX-Planck-Institut f$\ddot{\rm{u}}$r Extraterrestrische Physik, Giessenbachstrasse, D-85748, Garching, Germany
         \and Department of Physics, Technion-Israel Institute of Technology, 32000 Haifa, Israel
         \and Dipartimento di Matematica e Fisica, Universit{\`{a}} degli Studi Roma Tre, Via della Vasca Navale 84, 00146 Roma, Italy
         \and Mullard Space Science Laboratory, University College London, Holmbury St. Mary, Dorking, Surrey RH5 6NT, UK
         \and Departament de F{\'{i}}sica, EEBE, Universitat Polit{\`{e}}cnica de Catalunya, Av. Eduard Maristany 16, E-08019 Barcelona, Spain
         \and Univ. Grenoble Alpes, CNRS, IPAG, 38000 Grenoble, France
         \and Telespazio UK for the European Space Agency (ESA), European Space Astronomy Centre (ESAC), Camino Bajo del Castillo, s/n, E-28692 Villanueva de la Ca\~nada, Madrid, Spain
         \and Institute of Astronomy, University of Cambridge, Madingley Road, Cambridge CB3 0HA UK
         \and School of Physics and Astronomy and Wise Observatory, Tel Aviv University, Tel Aviv 69978, Israel
         \and Department of Astronomy, University of Geneva, 16 Ch. d0Ecogia, 1290 Versoix, Switzerland
         \and Italian Space Agency (ASI), Via del Politecnico snc, 00133 Roma, Italy
             }

  \abstract
  {
  The relation between warm absorber (WA) outflows of AGN 
  and nuclear obscuration activities caused by optically-thick 
  clouds (obscurers) crossing the line of sight is still unclear.
  NGC~3227 is a suitable target to study the 
  properties of both WAs and obscurers, 
  because it matches the following selection criteria:
   WAs in both ultraviolet (UV) and X-rays, suitably variable, bright in UV and X-rays,
   good archival spectra for comparing with the obscured spectra.
  To investigate WAs and obscurers of NGC~3227 in detail,
  we used a broadband spectral-energy-distribution model that is built in the first paper of our series 
  and the photoionization code of \texttt{SPEX} software to fit the archival observational data
  taken by \emph{XMM-Newton} and \emph{NuSTAR} in 2006 and 2016.
  Using unobscured observations, we find four WA components with different ionization 
  states ($\log\ \xi\ [{\rm{erg\ cm}}\ {\rm{s}}^{-1}] \sim -1.0,\ 2.0,\ 2.5,\ 3.0$).
  The highest-ionization WA component has a much higher hydrogen column 
  density ($\sim 10^{22}\ {\rm{cm}}^{-2}$) than the other three components ($\sim 10^{21}\ {\rm{cm}}^{-2}$). 
  The outflow velocities of these WAs range from 100 to 1300 ${\rm{km}}\ {\rm{s}}^{-1}$,
  and show a positive correlation with the ionization parameter.
  These WA components are estimated to be distributed from the outer region of the broad line region (BLR)
  to the narrow line region. 
  It is worth noting that we find an X-ray obscuration event in the beginning of the 2006 observation, 
  which was missed by previous studies. 
  It can be explained by a single obscurer component.
  We also study the previously published obscuration event captured in one observation in 2016,
  which needs two obscurer components to fit the spectrum.
  A high-ionization obscurer 
  component ($\log\ \xi \sim 2.80$; covering factor $C_{\rm{f}} \sim 30\%$) only appears in the 2016 observation, 
  which has a high column density ($\sim 10^{23}\ \rm{cm}^{-2}$). 
  A low-ionization obscurer component ($\log\ \xi \sim 1.0-1.9$; $C_{\rm{f}} \sim 20\%-50\%$)
  exists in both 2006 and 2016 observations,
  which has a lower column density ($\sim 10^{22}\ \rm{cm}^{-2}$). 
  These obscurer components are estimated to reside within the BLR by their crossing time of transverse motions.
  The obscurers of NGC~3227 are closer to 
  the center and have larger number densities than the WAs,
  which indicate that the WAs and obscurers might have different origins.
  }

   \authorrunning{Y.J. Wang et al.}
   \titlerunning{Warm absorbers and obscuration events in NGC 3227 seen with archival data}
   \maketitle

\section{Introduction}
\label{introd}

    Active galactic nuclei (AGN) accrete matter onto a central supermassive black hole (SMBH)
    to produce intense broadband radiation, which can ionize and drive away the surrounding matter in forms of outflows. 
    Many observational proofs have implied that outflows might play an important role in affecting the 
    star formation and evolution of their host galaxies \citep[see the review of][]{King2015}.
    Ionized outflows can be detected via absorption features along the line of sight in the ultraviolet (UV) and X-rays,
    which usually have different types \citep[][and references therein]{Laha2021} such as 
    broad absorption lines \citep[BALs;][]{Weymann1981},  
    warm absorbers \citep[WAs;][]{Halpern1984,Crenshaw2003}, and ultrafast outflows \citep[UFOs;][]{Tombesi2010}. 
    UFOs might have an origin close to the central engine \citep[$\sim 0.0003-0.03\ \rm{pc}$;][]{Tombesi2012} 
    with very high velocities \citep[$\sim 0.03-0.3c$;][]{Tombesi2010,Tombesi2012}.
    BALs usually reside outside the broad line region (BLR) with high outflow 
    velocities reaching $\sim 30000\ \rm{km}\ \rm{s}^{-1}$ \citep{Trump2006,Gibson2009}.
   Compared with UFOs and BALs, WAs have lower outflow velocities from about one hundred to a few 
   thousands of ${\rm{km}}\ {\rm{s}}^{-1}$ \citep[][]{Kaastra2000,Ebrero2013}, and they might 
   originate in the accretion disk \citep[e.g.,][]{Elvis2000,Krongold2007}, BLR 
   \citep[][]{Reynolds1995}, or dusty torus \citep[e.g.,][]{Krolik2001,Blustin2005}.
   Although different types of outflows have overlaps in distance scales and outflow parameters,
   the direct connection between these outflows still remains unclear.
   In this work, we mainly focus on the properties of the WA outflows.

   According to \cite{Tarter1969}, the ionization parameter can be defined by
   \begin{equation}
   \xi = \frac{L_{\rm{ion}}}{n_{\rm{H}} r^2},
   \label{equ:xi}
   \end{equation}
   where $L_{\rm{ion}}$ is the ionizing luminosity over 1--1000 Ryd, 
   $n_{\rm{H}}$ is the hydrogen number density 
   of the absorbing gas, and $r$ is the radial distance of the absorbing gas to 
   the central engine. 
   WAs might be driven by radiation pressure \citep[e.g.,][]{Proga2004}, magnetic 
   forces \citep[e.g.,][]{Blandford1982, Konigl1994, Fukumura2010}, 
   or thermal pressure \citep[e.g.,][]{Begelman1983, Krolik1995, Mizumoto2019},
   and show a wide range of 
   ionization parameter ($10^{-1}\leq \xi \leq 10^3\ {\rm{erg}}\ {\rm{cm}}\ {\rm{s}}^{-1}$) 
   and hydrogen column density ($10^{20}\leq N_{\rm{H}} \leq 10^{23}\ {\rm{cm}}^{-2}$) \citep{Laha2014}. 
   Investigating properties of WAs can help us to understand the formation of AGN outflows
   and their feedback efficiency to the host galaxy.
   These WAs have been found in about 50\% of nearby 
   AGN \citep[e.g.,][]{Reynolds1997,Kaastra2000,Porquet2004,Tombesi2013,Laha2014},
   and the properties of WAs show differences among different 
   AGN, such as the different ionization states,
   column densities, and outflow velocities \citep{Tombesi2013,Laha2014}.

   Moreover, the X-ray spectra of some AGN present dramatic hardening 
   accompanied by flux-drops on short timescales,
   which might be due to the X-ray transient obscuration events \citep{Markowitz2014}.
   Transient obscuration events can also cause 
   absorption features in the soft X-ray and UV bands,
   which usually appear and disappear on shorter timescales compared with outflows.
   These obscuration events might be explained by discrete optically thick clouds or gas clumps crossing 
   the line of sight, which are referred to as obscurers. These shielding gas clumps or obscurers might 
   ensure that the radiatively driven disk winds in broad absorption line quasars are not 
   over-ionized by UV/X-ray ionizing radiation and are accelerated further
   \citep{Murray1995,Proga2000,Kaastra2014}. 
   The obscuration events may be triggered by the collapse of 
   the BLR \citep{Kriss2019a,Kriss2019b,Devereux2021}.
   When the continuum radiation decreases, the BLR clouds will collapse towards
   the accretion disk; when the continuum brightens again, these collapsed clouds
   might be blown away as obscurers \citep{Kriss2019b}.
   X-ray obscuration events also have been found 
   in many AGN, such as NGC~5548 \citep{Kaastra2014}, 
   NGC~3783 \citep{Mehdipour2017,Kaastra2018,Marco2020}, 
   NGC~985 \citep{Ebrero2016a}, 
   and NGC~1365 \citep{Risaliti2007,Walton2014,Rivers2015}. 
   These obscurers may be located within 
   the BLR \citep{Lamer2003, Risaliti2007, Lohfink2012, Longinotti2013, Marco2020, Kara2021},
   or close to the outer BLR \citep[e.g.,][]{Kaastra2014, Beuchert2015, Mehdipour2017}, 
   or near the inner torus \citep[e.g.,][]{Beuchert2017}.

   Until now the relation between the WA outflows and the nuclear
   obscuration activity is not yet well understood. Whether the WAs and obscurers
   have the same origin or how shielding by the obscuration
   affects the WAs and their appearance is not well known. 
   Studying WAs in targets that have transient obscuration
   enables us to probe these questions.
   Transient obscuration events have been studied simultaneously in UV and X-rays
   in only a few AGN that have WAs outflows, 
   such as NGC~5548 \citep{Kaastra2014}, 
   and NGC~3783 \citep{Mehdipour2017}, 
   and also Mrk~335 \citep{Longinotti2013,Parker2019}. 
   Studying NGC~3227 (a Seyfert 1.5 galaxy at the redshift of 
   0.003859\footnote{The redshift of NGC~3227 is obtained from the
   NASA/IPAC Extragalactic Database
   (\href{https://ned.ipac.caltech.edu/}{NED}). The NED is funded by the National Aeronautics and Space Administration and operated by the California Institute of Technology.}) is a rare opportunity towards a 
   more general characterization. 
   NGC~3227 was one of eight suitable targets selected for the 
{\it{Neil Gehrels Swift Observatory}} monitoring/triggering
   programme \citep{Mehdipour2017}, which matches the following selection criteria:
   WAs in both UV and X-rays, suitably variable, bright in UV and X-rays,
   good archival spectra for comparing with the obscured spectra.
   Using our target of opportunity (ToO) monitoring programme
   of the {\it{Neil Gehrels Swift Observatory}},
   we captured another X-ray obscuration event in NGC~3227 
   in 2019 \citep[][hereafter Paper I]{Mehdipour2021}, 
   which was observed simultaneously with {\it XMM-Newton}, {\it NuSTAR}, 
   and Hubble Space Telescope/Cosmic Origins Spectrograph (HST/COS) to get a deeper 
   multi-wavelength understanding of the transient obscuration phenomenon in AGN.
   The studies of WA and the obscurer are interlinked, 
   so without having a proper model for the WA, 
   the new obscurer cannot be accurately studied.
   In this work (the second paper of our series), we aim to study a 
   comprehensive model for the WA, and then use this WA model to 
   investigate the obscuration events appearing in NGC~3227.

  It should be noted that photoionization modelling strongly depends on the 
  ionizing spectral-energy-distribution (SED). 
  Therefore, to properly derive the ionization structure of the WA, 
  having an accurate broadband SED model is important.
  A few papers have reported studies of the WAs in NGC~3227
  \citep{Komossa1997, Beuchert2015, Turner2018, Newman2021} and 
  its nuclear obscurations activities 
  \citep{Lamer2003, Markowitz2014, Beuchert2015, Turner2018}.
  However, the contribution of the SED components that dominate in the UV/optical band
  lacks consideration, which might affect the fitting 
  results of the WAs and obscurers (see Paper I).
  The main effect of using different SEDs is that the derived 
  ionization parameter $\xi$ would be different. The total hydrogen column density $N_{\rm{H}}$ of the 
  WA (i.e. sum of the individual components) would be similar, 
  but how $N_{\rm{H}}$ is distributed over different ionization components 
  depends on the SED. 
  We refer the readers to \cite{Mehdipour2016} where the effect of 
  using different SEDs, and codes, is shown.
  With these considerations, we firstly built a broadband SED model 
  from the near infrared (NIR) to hard X-rays for NGC~3227 in our Paper I. 
   In this paper, we use this broadband SED model 
   and a robust photoionization code \citep[\texttt{pion} model; ][]{Mehdipour2016} 
   in the {\texttt{SPEX}} package 
   \citep{Kaastra1996} v3.05.00  \citep{kaastra2020_4384188}
   to analyze the archival {\it XMM-Newton} and {\it NuSTAR} data 
   taken in 2006 \citep{Markowitz2009} and 2016 \citep{Turner2018}.
   \texttt{SPEX} is currently the only code that enables the 
   SED and the ionization balance to be fitted simultaneously, 
   while all other codes have to pre-calculate the ionization balance on a given SED.
   The \texttt{pion} model is a self-consistent model that can simultaneously 
   calculate the thermal/ionization balance and 
   the plasma spectrum in photoionization equilibrium.
   In this work, we focus on properties of the WAs and obscuration 
   events of NGC~3227 with the archival 2006 and 2016 data.
   The detailed analysis of the 2019 obscuration events will be presented in Paper III by \cite{Maoprep}.
   The discussion about how the obscurer changes over the course of the 2019 observations 
   with the XMM-Newton/EPIC-pn data will be presented in Paper IV by \cite{Samprep}. 

   This paper is organized as follows. In Sect. \ref{sec:observation}, we present 
   the archival data that are used in this work and the data reduction process. In Sect. \ref{sec:spectra}, 
   we introduce the spectral analysis based on the broadband SED model. In Sect. \ref{sec:results}, 
   we present and discuss the results about WAs and obscurer components. 
   In Sect. \ref{sec:summary}, we give a summary of our conclusions.
   In this work, Cash statistic \cite[][hereafter C-stat]{Kaastra2017} will be used to estimate the goodness of fit and 
   statistical errors will be given at 1$\sigma$ (68\%) confidence level. 
   We adopt the following flat $\Lambda$CDM cosmological parameters: 
   $\textit{H}_\textrm{0}$=70 km s$^{\textrm{-1}}$ Mpc$^{\textrm{-1}}$, 
   $\textrm{$\Omega$}_\textit{m}$=0.30, and $\textrm{$\Omega$}_{\textrm{$\Lambda$}}$=0.70.

   \begin{table}
   \caption{Archival {\xmm} and {\nustar} data used for spectral analysis. \label{table:table1}}
   \centering
   \setlength{\tabcolsep}{3pt}
   \begin{tabular}{cccccc}
   \hline\hline\xrowht[()]{10pt}
   \# & Observatory & ObsID & Date & Exposure \\
       &  &  & (yyyy-mm-dd) & (ks) \\
   \hline\xrowht[()]{10pt}
   Obs1 & \it{XMM}  & 0400270101   & 2006-12-03 & 108 \\
   \hline\xrowht[()]{10pt}
   Obs2 & \it{XMM} & 0782520201   & 2016-11-09 &   92 \\
               & \it{NuSTAR}         & 60202002002 & 2016-11-09 &  50 \\
   \hline\xrowht[()]{10pt}
   Obs3 & \it{XMM} & 0782520301   & 2016-11-25 & 74 \\
               & \it{NuSTAR}         & 60202002004 & 2016-11-25 & 43 \\
   \hline\xrowht[()]{10pt}
   Obs4 & \it{XMM} & 0782520501    & 2016-12-01 & 87 \\
               & \it{NuSTAR}         & 60202002008 & 2016-12-01 & 42 \\
   \hline\xrowht[()]{10pt}
   Obs5  & \it{XMM} & 0782520601 & 2016-12-05 & 87 \\
                & \it{NuSTAR}         & 60202002010 & 2016-12-05 & 41 \\
   \hline\xrowht[()]{10pt}
   Obs6  &\it{XMM}  & 0782520701 & 2016-12-09 & 88 \\
                & \it{NuSTAR}         & 60202002012 & 2016-12-09 & 39 \\
   \hline
   \end{tabular}
   \end{table}
 
\section{Observations and Data Reduction}
\label{sec:observation}

   In Table \ref{table:table1}, we list the archival data that are used in this work.
   These data include six {\xmm} observations \citep{Markowitz2009,Turner2018}
   and five {\nustar} observations \citep{Turner2018}. 
   We do not use the archival {\xmm} observations taken
   in 2000 and on 2016 November 29, because the spectrum of the 2000 observation
   has a lower signal to noise ratio (S/N) owing to its short exposure time, 
   and the observation on 2016 November 29 shows a relatively 
   unstable softness ratio curve \citep[see Fig. 1 of][]{Turner2018}, 
   which may bias the estimation of WAs parameters.

\subsection{XMM-Newton data}
\label{sec:xmm}

   The data reduction was done using {\xmm} Science Analysis 
   Software (SAS) version 18.0.0, following the standard data analysis 
   procedure\footnote{See \href{https://www.cosmos.esa.int/web/xmm-newton/sas-threads}
   {https://www.cosmos.esa.int/web/xmm-newton/sas-threads} for details.}. 
   The cleaned event files of EPIC-pn data were produced using the {\texttt{epproc}} 
   pipeline and flaring particle background larger than 0.4 count/s was excluded. 
   The EPIC-pn spectra and lightcurves were extracted from a circular region with 
   a radius of 30 arcsec for the source and from a nearby source-free circular region with 
   a radius of 35 arcsec for the background. Response matrices and ancillary response 
   files of each observation were produced using the SAS tasks {\texttt{arfgen}} and {\texttt{rmfgen}}. 
   Following the standard procedure, the first-order data of RGS1 and RGS2 were 
   extracted using the SAS task {\texttt{rgsproc}} and 
   flaring particle background larger than 0.2 count/s was excluded. 
   We then combined the spectra of RGS1 
   and RGS2 using the SAS task {\texttt{rgscombine}}. We refer readers to our Paper I
   for the detailed data reduction of the Optical Monitor (OM) data. 
   Only the OM UVW1 filter is available for both 2006 and 2016 observations.

\subsection{NuSTAR data}
\label{sec:xmm}

   For the two telescope modules (FPMA and FPMB) data of {\nustar}, 
   level 1 calibrated and level 2 cleaned event files were produced 
   using the standard procedure of the {\texttt{nupipeline}} task of HEASoft v6.27. The level 3 products 
   including lightcurves, spectra and response files were extracted using 
   the task {\texttt{nuproducts}} from a circular region with a radius of 90 arcsec 
   for the source and from a nearby source-free circular region with the same radius 
   for the background. Finally, we produced combined spectra of 
   FPMA and FPMB data using tasks {\texttt{mathpha}} and produced combined response 
   files using tasks {\texttt{addrmf}} and {\texttt{addarf}}. 

   \begin{figure*}[t!]
   \centering
   \includegraphics[width=0.9\linewidth, clip]{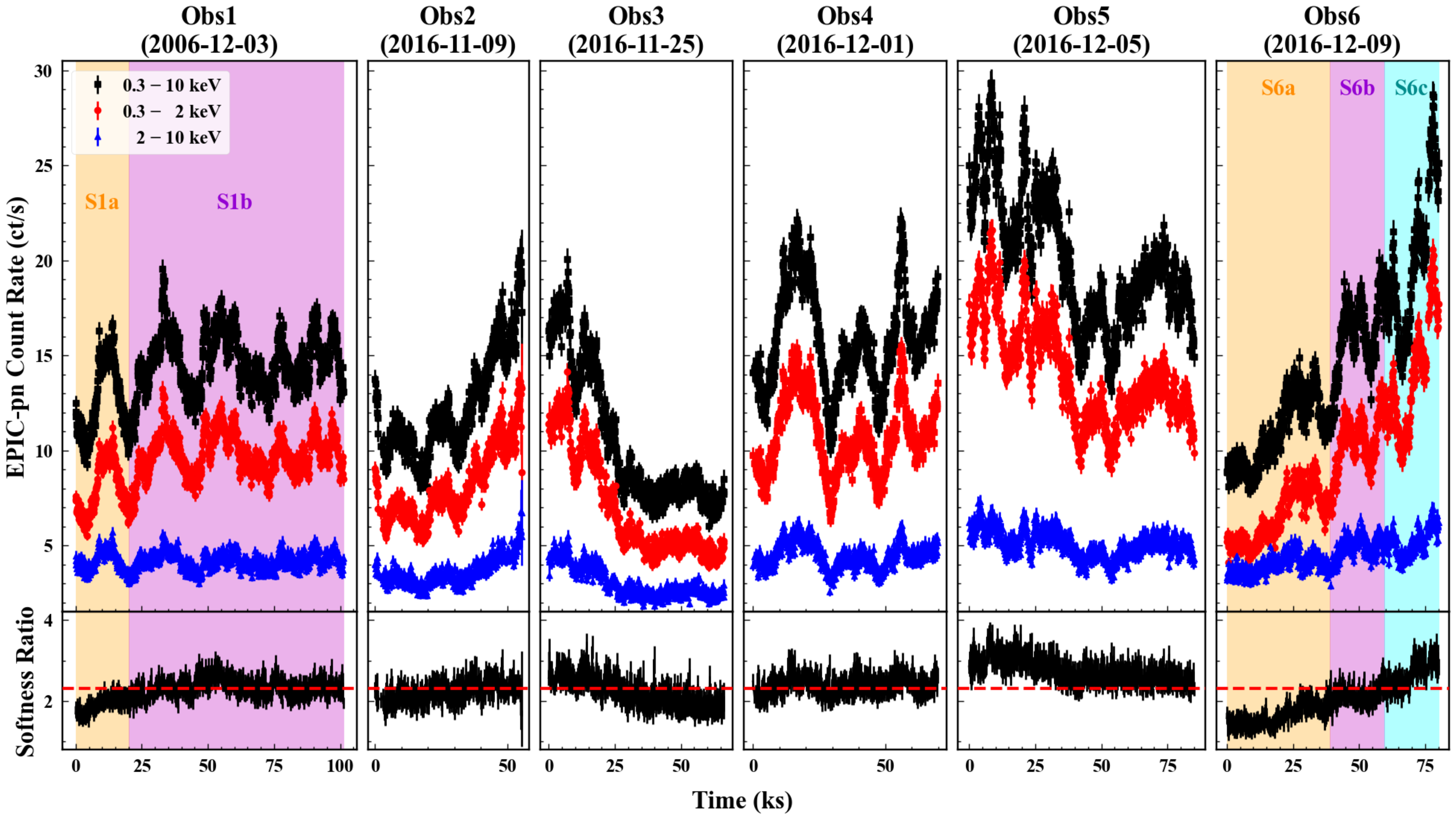}
   \caption{{\xmm}/EPIC-pn light curves (red points: 0.3--2 keV; 
   blue points: 2--10 keV; black points: 0.3--10 keV) of NGC~3227 ({\it top panel}) and 
   softness ratio curves between the count rates in 0.3--2 keV and 2--10 keV bands 
   ($\rm{Soft}_{\rm{0.3-2keV}}/\rm{Hard}_{\rm{2-10keV}}$; {\it bottom panel}) with the time bin of 100 s.
   The red horizontal dashed line in the bottom panel is the average softness ratio of all the six observations.
   According to the difference between the softness ratio of 
   each observation (black points in the bottom panel)
   and the average softness ratio (horizontal dashed line in red), 
   Obs1 is divided into two slices that are 
   S1a (orange region of the first column) and S1b (violet region of the first column), and
   Obs6 is divided into three slices that are S6a (orange region of the last column), 
   S6b (violet region of the last column), and S6c (cyan region of the last column).}
   \label{fig:lc}
   \end{figure*}

\begin{figure*}[t!]
\centering
\includegraphics[width=0.9\linewidth, clip]{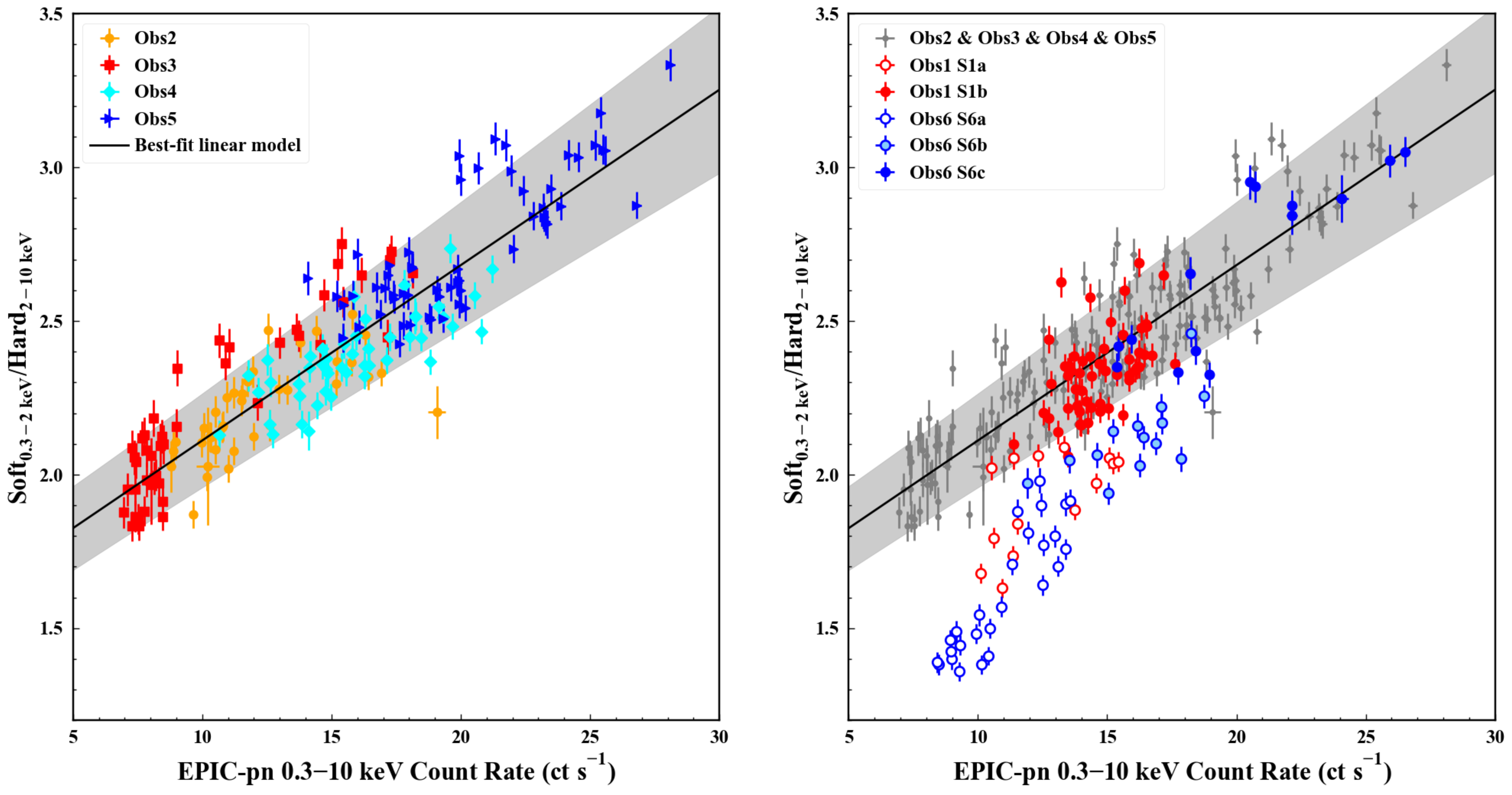}
\caption{Correlation between the softness ratio 
(the ratio of count rates between 0.3--2 and 2--10 keV bands) and 0.3--10 keV count rate,
for Obs2 to Obs5 ({\it left panel}), and for Obs1 and Obs6 ({\it right panel}). 
In both left and right panels, the black solid line is the best-fit linear model
of $y=(0.06\pm 0.01)  x +(1.54\pm 0.13)$ for Obs2 to Obs5 and the
shaded grey region is the associated 1$\sigma$ uncertainty.}
\label{fig:softness}
\end{figure*}

   \begin{table*}
   \caption{The best-fit parameters of the intrinsic broadband SED. \label{table:table2}}
   \centering
   \small
   \setlength{\tabcolsep}{3pt}
   \begin{tabular}{clcccccc}
   \hline\hline\xrowht[()]{10pt}
   Comp. & Parameter & Obs1 & Obs2 & Obs3 & Obs4 & Obs5 & Obs6 \\
   \hline\xrowht[()]{10pt}
  \texttt{dbb}    &   Normalization $A$ ($10^{26}\ \rm{cm}^2$)   &   5.87 (s)  &  7.51 (s)  &  7.15 (s)  &  6.87 (s) & 7.27 (s) & 7.39 (s)  \xrowht[()]{10pt} \\
  \hline\xrowht[()]{10pt}
  \texttt{comt} &   Normalization ($10^{54}\ \rm{ph}\ \rm{s}^{-1}\ \rm{keV}^{-1}$) & $1.70^{+0.24}_{-0.20}$ & $1.00^{+0.21}_{-0.25}$ & $0.50 \pm 0.15$ & $1.61^{+0.30}_{-0.27}$ & $1.06^{+0.05}_{-0.16}$ & $1.65^{+3.74}_{-0.19}$ \xrowht[()]{10pt} \\
  \hline\xrowht[()]{10pt}
  \texttt{pow} & Normalization ($10^{50}\ \rm{ph}\ \rm{s}^{-1}\ \rm{keV}^{-1}$) & $4.02 \pm 0.06$ & $2.93 \pm 0.06$ & $2.25 \pm 0.04$ & $3.60 \pm 0.06$ & $4.64^{+0.07}_{-0.01}$ & $4.59^{+0.25}_{-0.10}$ \xrowht[()]{10pt} \\
   & Photon index $\Gamma$ & $1.81 \pm 0.01$ & $1.73 \pm 0.02$ & $1.66 \pm 0.01$ & $1.78 \pm 0.009$ & $1.88 \pm 0.01$ & $1.86 \pm 0.03$ \xrowht[()]{10pt} \\
   \hline\xrowht[()]{10pt}
\texttt{refl} & Incident power-law Normalization & 4.02 (c) & 2.93 (c) & 2.25 (c) & 3.60 (c) & 4.64 (c) & 4.59 (c) \xrowht[()]{10pt} \\
   & Incident power-law photon index & 1.81 (c) & 1.73 (c) & 1.66 (c) & 1.78 (c) & 1.88 (c) & 1.86 (c) \xrowht[()]{10pt} \\
  & Reflection scale \emph{s} & $0.67 \pm 0.04$ & $0.54 \pm 0.07$ & $0.30 \pm 0.03$ & $0.42 \pm 0.03$ & $0.68 \pm 0.011$ & $0.57 \pm 0.05$ \xrowht[()]{10pt} \\
  \hline\xrowht[()]{10pt}
Luminosity                                                       & 0.3--2 keV luminosity $L_{\rm{0.3-2\ keV}}$ &  1.28  &  0.90  &  0.66  &  1.14  &  1.42  &  1.48      \xrowht[()]{10pt}      \\
 ($10^{42}\ {\rm{erg}}\ {\rm{s}}^{-1}$)               & 2--10 keV luminosity $L_{\rm{2-10\ keV}}$ &  1.45  &  1.20  &  1.00  &  1.32  &  1.50  &  1.54      \xrowht[()]{10pt}  \\
                                                           & 1--1000\ Ryd ionizing luminosity $L_{\rm{ion}}$ &  19.1  &  17.8  &  14.6  &  19.8  &  19.2  &   23.5      \xrowht[()]{10pt}   \\
                                                                                   & Bolometric luminosity $L_{\rm{bol}}$ &  43.3  & 47.1 & 42.4 &   46.6  &  47.1 &   52.2     \xrowht[()]{10pt}   \\
  \hline\xrowht[()]{10pt}
$C$-stat / $C$-expt.  &  (SED+WAs) &  3439$/$2794 (S1b)$^*$  &  4456$/$3621$^*$  &  3834$/$3584  &  4105$/$3588  & 4121$/$3570  &  3762$/$3638 (S6c) \xrowht[()]{10pt} \\
 (Best-fit) &      (SED+OC$_{\rm{L}}$+WAs)              &  3261$/$2926 (S1a)  &                         &                         &                         &                        &  3819$/$3670 (S6b) \xrowht[()]{10pt} \\
  & (SED+OC$_{\rm{H}}$+OC$_{\rm{L}}$+WAs)    &                         &                         &                         &                         &                        &  3786$/$3654 (S6a) \xrowht[()]{10pt} \\           
   \hline
   \end{tabular}
 \tablefoot{
The disk blackbody component (\texttt{dbb}) normalization
followed by ``(s)'' is obtained by scaling the \texttt{dbb} normalization of 
the 2019 observation (see Paper I) to match the OM UVW1 flux of each archival observation.
Given that the spectral shape of the emission from the outer disc
   might not have a strong variability on short timescales, for simplicity, 
   the following parameters are fixed to those of the 2019 observation:
   the \texttt{dbb} maximum temperature $T_{\rm{max}}=10\ {\rm{eV}}$,
   the warm Comptonized disk component (\texttt{comt}) seed photon temperature
   $T_{\rm{seed}}=10\ {\rm{eV}}$, the \texttt{comt} electron temperature 
   $T_{\rm{e}}=60\ {\rm{eV}}$, and the \texttt{comt} optical depth $\tau=30$.
   The normalization of the neutral reflection component (\texttt{refl})
   denoted by ``(c)'' is coupled to the normalization of the X-ray 
   power-law component (\texttt{pow}) in the fit, and the photon index ($\Gamma$) of the 
   \texttt{refl} denoted by ``(c)'' is coupled to the
   photon index of the \texttt{pow}.
   The 0.3--2 keV luminosity ($L_{\rm{0.3-2\ keV}}$),
   2--10 keV luminosity ($L_{\rm{2-10\ keV}}$),
   1--1000 Ryd ionizing luminosity ($L_{\rm{ion}}$),
   and bolometric luminosity ($L_{\rm{bol}}$)
   present the intrinsic luminosities. 
   The $C$-stat is the $C$-statistic value of the final 
   fitting result corresponding to the best-fit model 
   and the $C$-expt. is the expected $C$-stat.
   The best-fit model is shown as follows: 
   for Obs1 S1b, Obs2, Obs3, Obs4, Obs5, or Obs6 S6c, it is 
   the intrinsic SED plus the four warm absorbers (WAs), i.e., SED+WAs;
   for Obs S1a or Obs6 S6b, it is the intrinsic SED plus one low-ionization
   obscurer component (OC$_{\rm{L}}$) plus the four WAs, i.e., SED+OC$_{\rm{L}}$+WAs;
   for Obs6 S6a, it is the intrinsic SED plus one high-ionization obscurer component (OC$_{\rm{H}}$)
   plus OC$_{\rm{L}}$ plus the four WAs, i.e., SED+OC$_{\rm{H}}$+OC$_{\rm{L}}$+WAs.
   The best-fit SED parameters of Obs1 is obtained by fitting the spectra of S1b, 
   and the best-fit SED parameters of Obs6 is obtained by fitting the spectra of S6c.
   The SED parameters of S1a are fixed to those of S1b;
   the parameters of both S6a and S6b are fixed to those of S6c.
   Compared with other observations or slices, Obs1 S1b and Obs2 
   have relatively worse fitting results that are denoted by ``*'',
   which are mainly due to the uncertainties in the intercalibration 
   between {\it{XMM-Newton}} and {\it{NuSTAR}}.
   }
    \end{table*}

\section{Spectral Analysis}
\label{sec:spectra}

    For the 2016 archival data, we will consider each set of {\xmm} and {\nustar} observations 
   taken on the same date as a single dataset 
   (see Obs2 to Obs6 in Table \ref{table:table1}),
   where {\xmm} (OM UVW1 filter at $\sim$ 2910 $\AA$, 
   RGS data in the 6--37 $\rm{\AA}$ wavelength range, 
   and EPIC-pn data in the 2--10 keV energy band) and 
   {\nustar} (combined FPMA and FPMB data in the 5--78 keV energy range) data 
   are used simultaneously for the spectral analysis.
   For Obs1 that was taken on 2006 December 03, only {\xmm} observation is available (see Table \ref{table:table1}), 
   and we used Obs5 to verify that the best-fit parameters of the WAs do not significantly change without {\nustar} observation.
   In Figs. \ref{fig:lc} and \ref{fig:softness}, we show the {\xmm}/EPIC-pn light curves, softness ratio curves 
   (the ratio of count rates between 0.3--2 and 2--10 keV bands), and the correlation between 
   the softness ratio and 0.3--10 keV count rate for Obs1 to Obs6. According to
   these results, we make a preliminary analysis for the state of each observation.
   \begin{itemize}
   \item {\it{Obs1:}} This observation was considered to be in an 
   unobscured state by \cite{Markowitz2009}.
   However, compared with the average softness ratio of 
   the six observations (see the dashed red line in the bottom
   panel of Fig. \ref{fig:lc}),
   a significant spectral hardening (softness ratio below the 
   average value) occurred in the beginning of this observation
   (see the bottom panel of the first column in Fig. \ref{fig:lc}),
   which might indicate the existence of an obscuration event.
   After this spectral hardening period, the softness ratio of 
   Obs1 returns to the average value, 
   which might indicate that this obscuration event has disappeared. 
   These phenomena mean that these two periods of Obs1 should be analyzed separately.
   According to the difference between the softness ratio of 
   Obs1 and the average softness ratio of the six observations, 
   Obs1 is subdivided into the following two slices: S1a and S1b
   (see the first column in Fig. \ref{fig:lc}). 
   S1b is consistent with being in an unobscured state 
   as it follows the correlation of Obs2 to Obs5 
   (see the right panel of Fig. \ref{fig:softness}).
   However, S1a has a harder spectrum that is similar to 
   S6a and S6b (see the right panel of Fig. \ref{fig:softness}), 
   which might be in an obscured state, 
   but it was not reported in the previous works.
   \item {\it Obs2 to Obs5:} According to \cite{Turner2018}, Obs2 to Obs5 are in unobscured states,
   which show a linear correlation between the softness ratio (0.3--1$/$1--10 keV) and 0.3--10 keV count rate.
   We confirm this linear correlation for these observations 
   (softness ratio is calculated between the 0.3--2 and 2--10 keV bands in this work) in the left panel of Fig. \ref{fig:softness}.
   It is worth noting that although Obs2 and Obs3 show a spectral hardening during some periods, 
   these periods still follow the correlation of unobscured states (see the left panel of Fig. \ref{fig:softness}).
   Therefore, these periods might not be in obscured states.
   \item {\it Obs6:} \cite{Turner2018} had observed a rapid obscuration event in Obs6. To 
   restudy this obscuration event using a broadband SED model 
   and the \texttt{pion} model in {\texttt{SPEX}}, we followed \cite{Turner2018}
   to subdivide Obs6 into three slices: S6a, S6b, and S6c (see Fig. \ref{fig:lc}). 
   S6c shows an unobscured state as it follows the
   correlation of Obs2 to Obs5 (see the right panel of Fig. \ref{fig:softness}). 
   However, S6a and S6b deviate from the correlation of Obs2 to Obs5 (see the right panel of Fig. \ref{fig:softness}),
   therefore, might be in obscured states. 
   \end{itemize}  

   Next we make the detailed spectral analysis for these observational data.
   We begin our spectral modelling by using a broadband SED model from the NIR to 
   the hard X-ray bands for NGC~3227.
   We refer readers to our Paper I for full details about this SED model and we
   only give a brief introduction here.
   The main spectral components that are used in this work are shown below:
   \begin{enumerate}[1.]
   \item The intrinsic broadband SED (see details in Paper I), which 
   is composed of a disk blackbody component (\texttt{dbb}), a warm Comptonized disk component (\texttt{comt}) 
   from the optical to the soft X-ray band, an X-ray power-law component (\texttt{pow}), 
   and a neutral reflection component (\texttt{refl}) in the hard X-ray energy band. 
   For \texttt{pow} of NGC~3227, we used 309 keV \citep{Turner2018} as the high-energy exponential cut-off
   and used 13.6 eV as the low-energy exponential cut-off. 
   \item The obscurer components, which heavily absorb the X-ray spectrum. 
   We used the \texttt{pion} model in {\texttt{SPEX}} to fit their absorption features in the spectrum. 
   \item The warm absorber (WA) components, which produce absorption features 
   in soft X-rays. We also used the \texttt{pion} model to fit these absorption features.
    \item The Galactic X-ray absorption, which was taken into account  
   by the \texttt{hot} model in {\texttt{SPEX}} with the hydrogen 
   column density $N_{\rm{H}}=2.07 \times 10^{20}\ {\rm{cm}}^{-2}$ \citep{Murphy1996}.
   \end{enumerate}

   Obs2 to Obs5, S1b, and S6c are in unobscured states, 
   so we will fit their spectra using the spectral components 1, 3, and 4. 
   S1a, S6a, and S6b are in obscured states, 
   so their spectra will be fitted with the spectral components 1--4.
   Next we will present the details of the spectral analysis for spectral components 1, 2, and 3.

\subsection{The intrinsic broadband SED}
\label{sec:sed}

   For the archival data, only the OM UVW1 filter data is available for all the six observations, 
   so the parameters of the \texttt{dbb} component might not be well constrained in the fit.
   Moreover we do not expect a strong variability in the shape of the emission 
   from the outer disc on short timescales and commonly the variability of the 
   flux in long wavelengths is significantly smaller than that in the X-ray band.
   Therefore, we assume that the shape of the \texttt{dbb} component of each 
   archival observation is similar to that of the 2019 observation which has 
   optical and UV observational data to constrain the \texttt{dbb} component (see Paper I).  
   That is to say, we fixed the maximum temperature $T_{\rm{max}}$ of 
   the \texttt{dbb} component of the archival data
   to the 10 eV of the 2019 observation (see Paper I),
   and scaled the \texttt{dbb} normalization of the archival observations according to the 
   normalization of the 2019 observation (the scale factor of each archival observation is 
   the OM UVW1 flux ratio between the 2019 observation and each archival observation).
   For the \texttt{comt} component, its normalization was free in the fit and 
   the following parameters were fixed to those of the 2019 observation (see details in Paper I):  
   seed photon temperature $T_{\rm{seed}}=10$ eV, electron temperature $T_{\rm{e}}=60$ eV, 
   and optical depth $\tau=30$. 
   The \texttt{dbb} and \texttt{comt} components mainly dominate in the energy 
   band below 0.5 keV (see Paper I), so fixing the shapes of these two components
   might bring uncertainties to the parameter estimates for the WAs. 
   Even so, the normalizations of these two components, which are free in the fits,
   are still the main factors to affect the fitting result.
   For each slice in an obscured state (S1a, 
   S6a, and S6b), we assumed that it has the same intrinsic broadband SED 
   as the unobscured slice in the same observation (S1b, S6c, and S6c, respectively). Therefore, 
   we fixed the parameters of \texttt{dbb}, \texttt{comt}, \texttt{pow}, and \texttt{refl} components of 
   S1a to those of S1b in the fit. Similarly, these 
   parameters of S6a and S6b were fixed to those of S6c. For the \texttt{refl} component, 
   the scaling factor of the reflected spectrum (\emph{s}) was free in the fit, the 
   incident power-law normalization and photon index were coupled to those of the \texttt{pow} component.
   We summarize the best-fit parameters of the intrinsic broadband SED in Table \ref{table:table2},
   which will be discussed in Section \ref{sec:SED}.

   \begin{figure}
   \centering
   \includegraphics[width=\linewidth, clip]{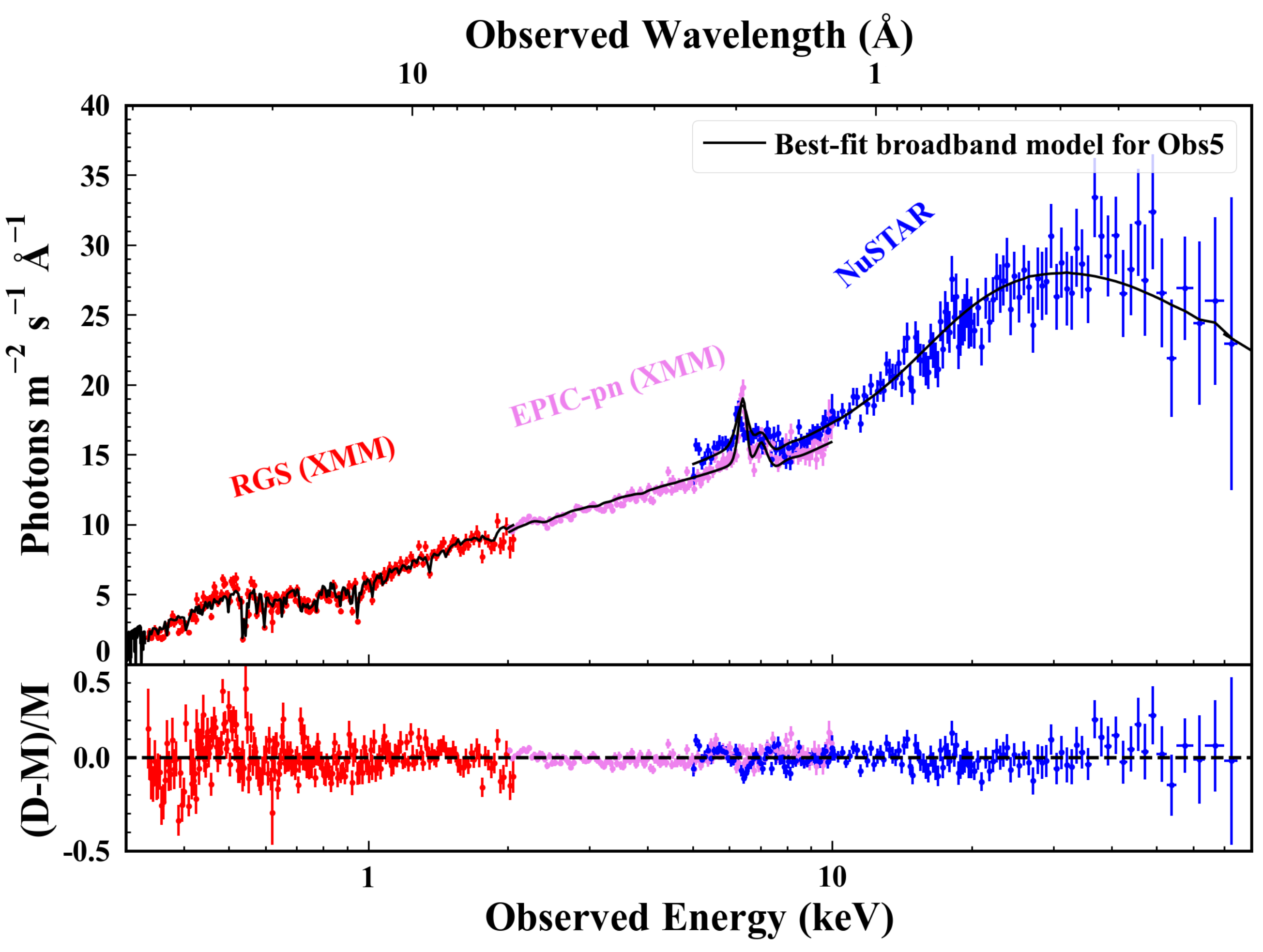}
   \caption{{\it Top panel:} The observational data (colored data points) with the 
   best-fit broadband SED model in the X-ray band (black solid curve) for Obs5.
   {\it Bottom panel:} Residuals of the best-fit model 
   (``D'' is the observational data and ``M'' is the best-fit model)
   for Obs5. Obs5 is taken as an example (other observations have similar fitting results). }
   \label{fig:datamodel}
   \end{figure}

\subsection{Warm absorber components}
\label{sec:waa}

   The hydrogen column density ($N_{\rm{H}}$), outflow velocity ($\varv_{\rm{out}}$) 
   and turbulent velocity ($\sigma_{\varv}$) of the WAs 
   are not well constrained simultaneously for the spectrum with low S/N and these parameters 
   might not vary significantly between different observations.
   Therefore, we fixed these parameters of Obs1--Obs4 and 
   Obs6 to the best-fit results of Obs5 (see Table \ref{table:table3}), 
   because Obs5 has the highest S/N. Therefore, for Obs1--Obs4 and Obs6, 
   only $\xi$ was free in the fit (see Table \ref{table:table3}). 
   Actually, $N_{\rm{H}}$, 
   $\varv_{\rm{out}}$, and $\sigma_{\varv}$ are not expected to be constant, 
   so our assumption will bring extra uncertainties to the parameter 
   estimates. However, $N_{\rm{H}}$, 
   $\varv_{\rm{out}}$, and $\sigma_{\varv}$ might not vary significantly on short
   timescales compared with $\xi$, so a significant impact on the fitting results with
   these parameters fixed might not be expected.
   For simplicity, we assumed 
   that the WAs fully cover the X-ray source (covering factor $C_{\rm{f}}=1$) and 
   have solar abundances \citep{Lodders2009}.
   The best-fit results of the WAs are discussed in Sect. \ref{sec:wa}.

\subsection{Obscurer components}
\label{sec:occ}

   Parameters $\varv_{\rm{out}}$ and $\sigma_{\varv}$ of the obscurer components were difficult to 
   constrain owing to the lack of well defined and strong absorption lines, 
   and we verified that changing their values had little impact on other parameters. 
   Therefore, we fixed $\varv_{\rm{out}}$ and $\sigma_{\varv}$ of the obscurer components to their default values 
   ($\varv_{\rm{out}} = 0\ \rm{km}\ \rm{s}^{-1}$ and $\sigma_{\varv}=100\ \rm{km}\ \rm{s}^{-1}$). 
   With that, the obscurer components, $N_{\rm{H}}$, $\xi$, and $C_{\rm{f}}$ are free in the fit.
   We will discuss the best-fit results of obscurer components in Sect. \ref{sec:obscurer}.

\section{Results and Discussion}
\label{sec:results}

\subsection{Intrinsic broadband SED}
\label{sec:SED}

\begin{figure}
   \centering
   \includegraphics[width=\linewidth, clip]{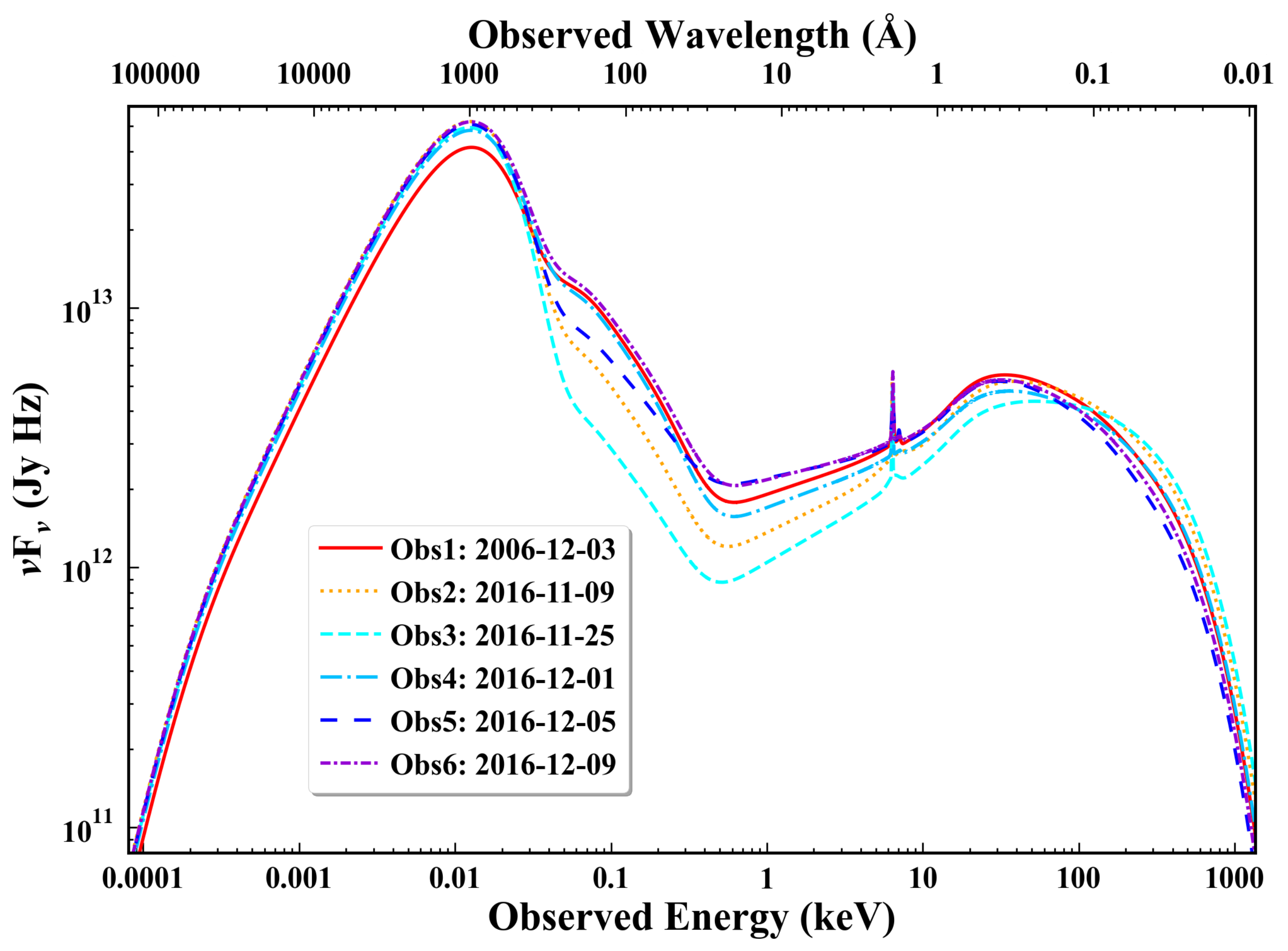}
   \caption{Intrinsic unabsorbed broadband SEDs of Obs1--Obs6.}
   \label{fig:sed}
   \end{figure}

   The broadband SED model provides a good description for the observational data 
   (see Table \ref{table:table2} and Fig. \ref{fig:datamodel}: Obs5 is taken as an example).
   Compared with other observations or slices, Obs1 S1b and Obs2 have relatively
   worse fitting results (see Table \ref{table:table3}),
   which are mainly due to the uncertainties in the intercalibration between {\it XMM-Newton} and {\it NuSTAR}.
   The intrinsic unabsorbed broadband SEDs of Obs1 to Obs6 are shown in Fig. \ref{fig:sed} 
   and their best-fit parameters are listed in Table \ref{table:table2}. The intrinsic broadband 
   SED of NGC~3227 shows a significant variability at energies $\geq$ 0.03 keV especially in the X-ray band
   (see Fig. \ref{fig:sed}). 
   According to the Spearman’s rank method, there is a positive correlation between the 
   photon index of the \texttt{pow} component ($\Gamma$; see Table \ref{table:table2}) and 
   the intrinsic 2--10 keV luminosity ($L_{\rm{2-10\ keV}}$; see Table \ref{table:table2})
   (the correlation coefficients $r_{\rm{s}}=0.94$ and the associated $p$-values $p_{\rm{s}}=0.005$), 
   which shows a ``softer-when-brighter'' behavior. This behavior 
   has been observed in many Seyfert galaxies \citep[e.g.,][]{Markowitz2004,Ponti2006,Sobolewska2009,Soldi2014}. 
   \cite{Peretz2018} also found a softer-when-brighter variability behavior in NGC~3227, which might be driven by 
   varying absorption rather than by the intrinsic variability of the central source.
   The averaged ionizing luminosity over 1--1000 Ryd is around 
   $1.9\times 10^{43}\ {\rm{erg}}\ {\rm{s}}^{-1}$ (see Table \ref{table:table2}), 
   which is two times larger than the results of \cite{Beuchert2015} and \cite{Turner2018}.
   This might be due to the different SED model in the UV and soft X-ray bands.
   The averaged full-band bolometric luminosity ($L_{\rm{bol}}$) is about 
   $4.6\times 10^{43}\ {\rm{erg}}\ {\rm{s}}^{-1}$ (see Table \ref{table:table2}), 
   which is 1.5 times smaller than the result of \cite{Woo2002} based on the flux integration method. 
   According to the reverberation mapping method, 
   the black hole mass ($M_{\rm{BH}}$) of NGC~3227 is 
   $5.96\times 10^6\ M_{\odot}$ for NGC~3227 \citep{Bentz2015}.
   The Eddington luminosity ($L_{\rm{Edd}}$) of NGC~3227 is $7.45\times 10^{44}\ \rm{erg}\ \rm{s}^{-1}$,
   which is calculated by $L_{\rm{Edd}}=1.25\times 10^{38}\times (M_{\rm{BH}}/M_{\odot})$
   \citep{Rybicki1979}. Therefore, the averaged Eddington ratio of NGC~3227 is about 6\%.

   \begin{table*}
   \caption{The best-fit parameters of the 
   four WAs (WA$_1$, WA$_2$, WA$_3$, WA$_4$) for Obs1 to Obs6: 
   hydrogen column density ($N_{\rm{H}}$), ionization parameter ($\xi$), 
   outflow velocity ($v_{\rm{out}}$), and turbulent velocity ($\sigma_v$). \label{table:table3}}
   \centering
   \tiny
   \setlength{\tabcolsep}{4pt}
   \begin{tabular}{cl|cc|cccc|ccc}
   \hline\hline\xrowht[()]{10pt}
      &  & \multicolumn{2}{c|}{03 Dec. 2006} & 09 Nov. 2016 & 25 Nov. 2016 & 01 Dec. 2016 & 05 Dec. 2016 & \multicolumn{3}{c}{09 Dec. 2016} \\
Component & Parameter & \multicolumn{2}{c|}{Obs1} & Obs2 & Obs3  & Obs4 & Obs5 & \multicolumn{3}{c}{Obs6} \\
\cline{3-4} \cline{9-11}
 &  & S1b & S1a$^*$ &  &  &  & & S6c & S6b$^*$ & S6a$^*$ \xrowht[()]{10pt} \\
   \hline\xrowht[()]{7pt}
         & $N_{\rm{H}}\ (10^{22}\ {\rm{cm}}^{-2})$ & 2.27 (f) & 2.27 (f) & 2.27 (f) & 2.27 (f) & 2.27 (f) & $2.27^{+0.08}_{-0.19}$ & 2.27 (f) & 2.27 (f) & 2.27 (f) \xrowht[()]{7pt} \\
WA$_1$ & $\log\ [\xi\ ({\rm{erg\ cm\ s}}^{-1})]$ & $2.96\pm 0.04$ & 2.93 & $2.92\pm 0.04$ & $3.21\pm 0.06$ & $3.29\pm 0.05$ & $3.29^{+0.08}_{-0.04}$  & $3.04^{+0.24}_{-0.06}$ & 3.02 & 2.94 \xrowht[()]{7pt} \\
         & $\varv_{\rm{out}}\ ({\rm{km\ s}}^{-1})$ & $-$1270 (f) & $-$1270 (f) & $-$1270 (f) & $-$1270 (f) & $-$1270 (f) & $-1270^{+20}_{-120}$ & $-$1270 (f) & $-$1270 (f) & $-$1270 (f) \xrowht[()]{7pt} \\ 
         & $\sigma_{\varv}\ ({\rm{km\ s}}^{-1})$ & 20 (f) & 20 (f) & 20 (f) & 20 (f) & 20 (f) & $20^{+20}_{-10}$ & 20 (f) & 20 (f) & 20 (f) \xrowht[()]{7pt} \\
         \hline
         & $N_{\rm{H}}\ (10^{22}\ {\rm{cm}}^{-2})$ & 0.25 (f) & 0.25 (f) & 0.25 (f) & 0.25 (f) & 0.25 (f) & $0.25\pm 0.04$  & 0.25 (f) & 0.25 (f) & 0.25 (f) \xrowht[()]{7pt} \\
WA$_2$ & $\log\ [\xi\ ({\rm{erg\ cm\ s}}^{-1})]$ & $2.54\pm 0.02$ & 2.51 & $2.68\pm 0.03$ & $2.74\pm 0.04$ & $2.65\pm 0.03$ & $2.59\pm 0.02$ & $2.58^{+0.21}_{-0.03}$ & 2.56 & 2.47 \xrowht[()]{7pt} \\
         & $\varv_{\rm{out}}\ ({\rm{km\ s}}^{-1})$ & $-$500 (f) & $-$500 (f) & $-$500 (f) & $-$500 (f) & $-$500 (f) & $-500^{+60}_{-50}$ & $-$500 (f) & $-$500 (f) & $-$500 (f) \xrowht[()]{7pt} \\
         & $\sigma_{\varv}\ ({\rm{km\ s}}^{-1})$ & 140 (f) & 140 (f) & 140 (f) & 140 (f) & 140 (f) & $140^{+10}_{-20}$ & 140 (f) & 140 (f) & 140 (f) \xrowht[()]{7pt} \\   
\hline
         & $N_{\rm{H}}\ (10^{22}\ {\rm{cm}}^{-2})$ & 0.12 (f) & 0.12 (f) & 0.12 (f) & 0.12 (f) & 0.12 (f) & $0.12\pm 0.012$  & 0.12 (f) & 0.12 (f) & 0.12 (f) \xrowht[()]{7pt}  \\
WA$_3$ & $\log\ [\xi\ ({\rm{erg\ cm\ s}}^{-1})]$ & $1.91\pm 0.07$ & 1.88 & $2.22\pm 0.09$ & $2.55\pm 0.13$ & $1.84\pm 0.08$ & $1.89\pm 0.07$ & $1.97^{+0.19}_{-0.15}$ & 1.94 & 1.86 \xrowht[()]{7pt} \\
         & $\varv_{\rm{out}}\ ({\rm{km\ s}}^{-1})$ & $-$440 (f) & $-$440 (f) & $-$440 (f) & $-$440 (f) & $-$440 (f) & $-440^{+160}_{-80}$ & $-$440 (f) & $-$440 (f) & $-$440 (f) \xrowht[()]{7pt} \\
         & $\sigma_{\varv}\ ({\rm{km\ s}}^{-1})$ & 50 (f) & 50 (f) & 50 (f) & 50 (f) & 50 (f) & $50 \pm 20$ & 50 (f) & 50 (f) & 50 (f) \xrowht[()]{7pt} \\    
\hline
         & $N_{\rm{H}}\ (10^{22}\ {\rm{cm}}^{-2})$ & 0.16 (f) & 0.16 (f) & 0.16 (f) & 0.16 (f) & 0.16 (f) & $0.16\pm 0.007$ & 0.16 (f) & 0.16 (f) & 0.16 (f) \xrowht[()]{7pt} \\
WA$_4$ & $\log\ [\xi\ ({\rm{erg\ cm\ s}}^{-1})]$ & $-1.25 \pm 0.11$ & $-1.28$ & $-1.13^{+0.20}_{-0.16}$ & $-0.69^{+0.17}_{-0.20}$ & $-1.40 \pm 0.16$ & $-1.06\pm 0.11$ & $-1.23^{+0.12}_{-0.98}$ & $-1.25$ & $-1.28$ \xrowht[()]{7pt} \\
         & $\varv_{\rm{out}}\ ({\rm{km\ s}}^{-1})$ & $-$110 (f) & $-$110 (f) & $-$110 (f) & $-$110 (f) & $-$110 (f) & $-110^{+90}_{-40}$ & $-$110 (f) & $-$110 (f) & $-$110 (f) \xrowht[()]{7pt} \\
         & $\sigma_{\varv}\ ({\rm{km\ s}}^{-1})$ & 260 (f) & 260 (f) & 260 (f) & 260 (f) & 260 (f) & $260^{+40}_{-30}$ & 260 (f) & 260 (f) & 260 (f) \xrowht[()]{7pt} \\    
   \hline
   \end{tabular}
   \tablefoot{The parameters followed by ``(f)'' are fixed to the value of Obs5 that has the highest S/N.
   The slices followed by ``*'' mean that they are in obscured states. 
   }
   \end{table*}

\subsection{Warm absorber components}
\label{sec:wa}

   At least four WA components (see Table \ref{table:table3}) were required to 
   improve the fitting result ($\Delta C \sim 100$ for adding WA$_1$, 
   $\Delta C \sim 80$ for adding WA$_2$, $\Delta C \sim 80$ for adding WA$_3$, $\Delta C \sim 500$ for adding WA$_4$). 
   We added another component (a fifth component),
   but it did not improve the fitting result ($\Delta C \sim 2$). 
   Our result is not consistent with that in \cite{Turner2018}, which detected three WA components.
   It might be due to the different SED model and photoionization models between our and their works.
   The best-fit model shows some weak residual emission features in the 0.5--0.6 keV band (see Fig. \ref{fig:datamodel})
   caused by the oxygen line emission from distant regions, similar to NGC~5548 \citep{Mao2018}.
   The quality of our spectrum is insufficient to do detailed modelling of these emission lines, and 
   they are too weak to affect our modelling of the absorbing wind components. 
   Therefore they will not be discussed further.

\subsubsection{Parameters of the warm absorbers}
\label{sec:parawa}

\begin{figure}
\centering
\includegraphics[width=\linewidth, clip]{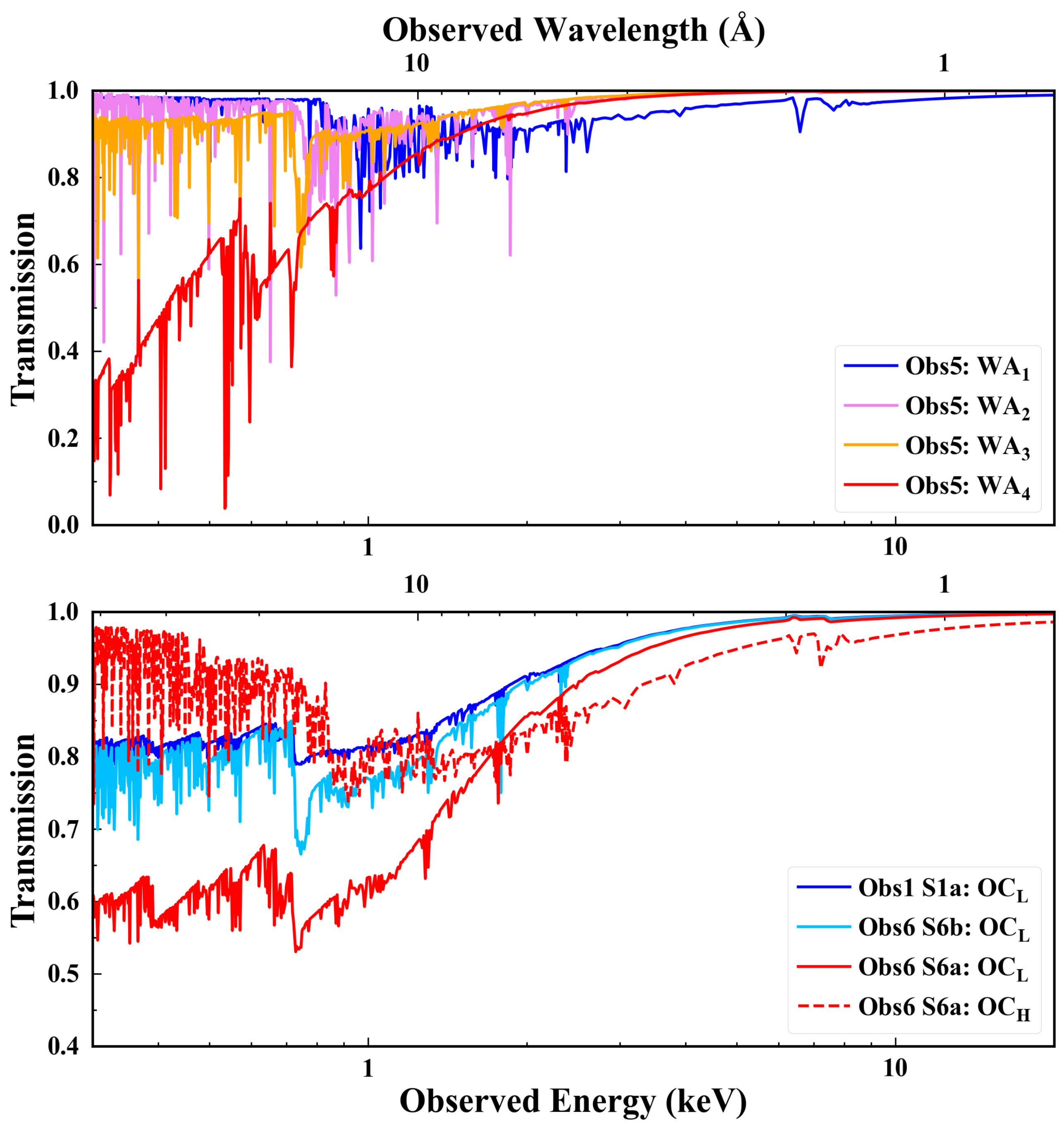}
\caption{Transmission spectra of the four WAs of Obs5 
(WA$_1$, WA$_2$, WA$_3$, WA$_4$; {\it top panel}; Obs5 is taken as an example) and 
obscurer components of S6a, S6b and S1a ({\it bottom panel}). OC$_{\rm{L}}$ is the low-ionization 
obscurer component while OC$_{\rm{H}}$ is the high-ionization obscurer component.}
\label{fig:transmission}
\end{figure}

   The logarithm of the ionization parameter (in units of ${\rm{erg\ cm\ s}}^{-1}$) 
   of the four WA components is around 
   3.0 for WA$_1$, 2.5 for WA$_2$, 2.0 for WA$_3$, and 
   $-$1.0 for WA$_4$ (see Table \ref{table:table3}). 
   WA$_2$, WA$_3$, and WA$_4$ have a similar column 
   density around $10^{21}\  \rm{cm}^{-2}$, while WA$_1$ has a much higher value near 
   $10^{22}\  \rm{cm}^{-2}$ (see Table \ref{table:table3}). 
   In paper I, we adopt a de-ionization scenario for the WAs and 
   simultaneous fitted the spectra of the archival unobscured observation taken on 
   2016 December 05 and the new obscured observations taken in 2019. 
   In this work, we fit the spectra taken on 2016 December 05 alone.
   Therefore, some different results between Paper I and Paper II 
   can be expected because of the different spectral modelling. 
   The X-ray transmission of each WA component in our line of sight to the central region
   is shown in the top panel of Fig. \ref{fig:transmission}.
   WA$_1$ mainly absorbs the continuum radiation between 0.8 and 10 keV. 
   WA$_2$ and WA$_3$ produce absorption features between 0.7 and 5 keV. 
   WA$_4$ heavily absorbs the continuum below 5 keV.
   From WA$_1$ to WA$_4$, the outflow velocity gradually decreases from $\sim 1300$ 
   to $\sim 100\ \rm{km}\ \rm{s}^{-1}$ (see Table \ref{table:table3}), which shows a positive 
   correlation with the ionization parameter.
   This correlation is consistent with the results for AGN samples \citep{Tombesi2013,Laha2014}.
   These previous studies indicated that this correlation cannot be explained by the 
   radiatively driven or magneto hydrodynamically 
   driven outflowing mechanism \citep{Tombesi2013,Laha2014},
   and it might be explained by the equilibrium between the radiation pressure on WAs and the 
   drag pressure from the ambient circumnuclear medium \citep[see details in][]{Wangsubmit}.

\subsubsection{Radial location of the warm absorbers}
\label{sec:locawa}

   We use three methods to estimate the upper or lower limit of the locations of the various WA components.
   First, we assume that the thickness ($\Delta r$) of the WA cloud
   does not exceed its distance ($r$) to the SMBH \citep{Krolik2001,Blustin2005}. 
   As $N_{\rm{H}} \approx n_{\rm{H}} C_{\rm{v}} \Delta r$, so the upper limit of the distance $r_{\rm{max}} = \Delta r = N_{\rm{H}}/(n_{\rm{H}} C_{\rm{v}})$,
   where $C_{\rm{v}}$ is the volume filling factor.
   Combining with Eq. \ref{equ:xi}, $r_{\rm{max}}$ can be estimated by
   \begin{equation}
   r_{\rm{max}} = \frac{L_{\rm{ion}} C_{\rm{v}}}{\xi N_{\rm{H}}}.
   \label{equ:rmax}
   \end{equation}
   Assuming that the total outflow momentum of the WA cloud is equal to the momentum of the absorbed radiation ($P_{\rm{abs}}$)
   plus the momentum of the ionizing luminosity being scattered ($P_{\rm{scat}}$),
   $C_{\rm{v}}$ can be calculated by
   \begin{equation}
   C_{\rm{v}} = \frac{(\dot{P}_{\rm{abs}}+\dot{P}_{\rm{scat}})\xi}{1.23 m_{\rm{p}} L_{\rm{ion}} v_{\rm{out}}^2 \Omega}
   \label{eq:cv}
   \end{equation}
   \citep{Blustin2005,Grafton2020}. Here, $\dot{P}_{\rm{abs}}$ is given by
   \begin{equation}
   \dot{P}_{\rm{abs}} = \frac{L_{\rm{abs}}}{c},
   \end{equation}
   where $L_{\rm{abs}}$ is the absorbed luminosity, and $\dot{P}_{\rm{scat}}$ is calculated by
   \begin{equation}
   \dot{P}_{\rm{scat}} = \frac{L_{\rm{ion}}}{c}(1-e^{-\tau_{\rm{T}}});\ \tau_{\rm{T}} = \sigma_{\rm{T}} N_{\rm{H}},
   \label{equ:pscat}
   \end{equation}
   where $\tau_{\rm{T}}$ is the optical depth for Thomson scattering, 
   and $\sigma_{\rm{T}}$ is the Thomson scattering cross-section.
   We use the ionizing luminosity of Obs5 to estimate the distance as
   the WAs parameters are mainly from the spectral fitting of Obs5.
   The estimated upper limit ($r_{\rm{max}}$ in Eq. \ref{equ:rmax}) 
   of the radial location of each WA component is summarized in Table \ref{table:table5},
   which is 0.007 pc for WA$_1$, 0.24 pc for WA$_2$, 0.71 pc for WA$_3$, and 265 pc for WA$_4$.

   The second method is based on the assumption that the outflow velocities of winds are larger than or equal to their escape 
   velocities $\varv_{\rm{esc}}=\sqrt{2GM_{\rm{BH}}/r}$ \citep{Blustin2005}, 
   then we can obtain the lower limit of $r$ by
   \begin{equation}
   r_{\rm{min}} = \frac{2GM_{\rm{BH}}}{\varv_{\rm{out}}^2},
   \label{equ:rmin}
   \end{equation}
   where $G$ is the gravitational constant. 
   The lower limit of the radial location of each WA component is estimated to be 
   0.03 pc for WA$_1$, 0.2 pc for WA$_2$, 0.3 pc for WA$_3$, 
   and 4 pc for WA$_4$ (see Table \ref{table:table5}).

\begin{figure*}[t!]
\centering
\includegraphics[width=0.9\linewidth, clip]{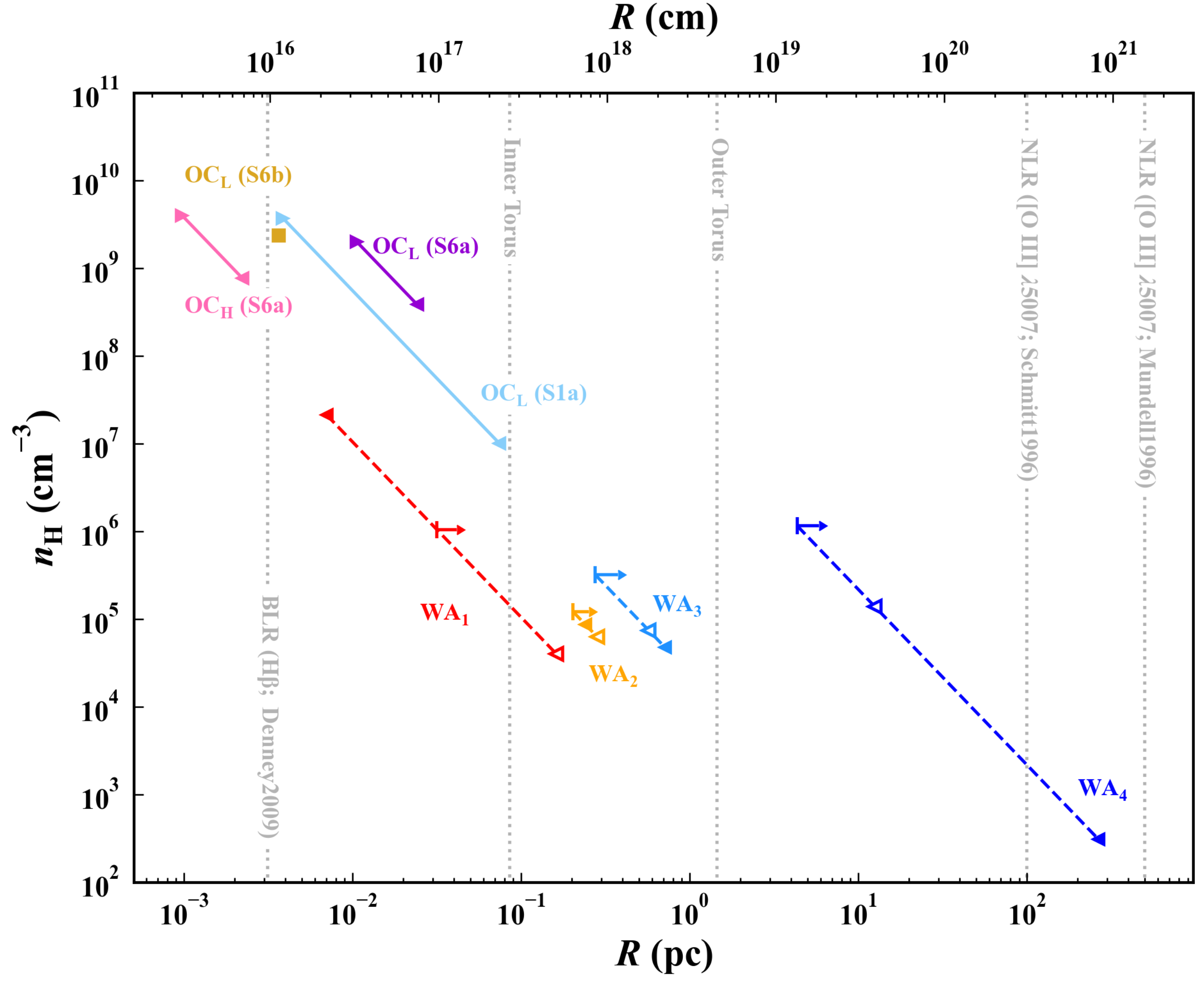}
\caption{Distances of the four WAs (WA$_1$, WA$_2$, WA$_3$, WA$_4$) 
and obscurer components from the center of NGC~3227.
OC$_{\rm{L}}$ is the low-ionization 
obscurer component and OC$_{\rm{H}}$ is the high-ionization obscurer component.
Symbols ``{\large{$\blacktriangleleft$}}'' for the WAs
represent the upper limits of the distance estimated 
by the assumption that the thickness of the WA cloud
does not exceed its distance to the SMBH (see Eq. \ref{equ:rmax}). 
Symbols ``$\mapsto$'' for the WAs
represent the lower limits of the distance estimated 
by the assumption that the outflow velocities of winds are larger than or equal to their escape velocities (see Eq. \ref{equ:rmin}). 
Symbols ``{\LARGE{$\triangleleft$}}'' for the 
WAs represent the upper limits of the distance estimated by the upper limits of 
the recombination timescale (see Eq. \ref{equ:trec}).
Symbols ``{\large{$\blacktriangleright$}}'' and ``{\large{$\blacktriangleleft$}}'' for 
the obscurer components represent the lower and upper limits of the distance estimated 
by the range of the crossing time (see Eq. \ref{equ:dist}). 
The radial location of the optical BLR of NGC~3227 is estimated by the time 
lag between the H$\beta$ line and continuum at 5100 $\AA$ \citep{Denney2009}. 
The inner radii of the torus is estimated by the dust sublimation radius (see Eq. \ref{equ:torusin})
and its outer radii is estimated by $R_{\rm{out}}=Y\times R_{\rm{in}}$ \citep[$Y=17$ for NGC~3227;][]{Alonso2011}.
The location of the NLR is estimated by the optical [O III] $\lambda$5007 image 
\citep{Mundell1995,Schmitt1996}.}
\label{fig:distance}
\end{figure*}

   The third method to estimate the radial location of WAs is based on the 
   recombination timescale. Following \cite{Bottorff2000}, the recombination timescale of the ion $X_i$ is defined as
   \begin{equation}
   t_{\rm{rec}}(X_i) = \bigg( \alpha_{\rm{r}}(X_i) n_{\rm e} 
   \bigg[\frac{f(X_{i+1})}{f(X_i)} - \frac{\alpha_{\rm{r}}(X_{i-1})}{\alpha_{\rm{r}}(X_i)} \bigg]\bigg)^{-1},
   \label{equ:trec}
   \end{equation}
   where $\alpha_{\rm{r}}(X_i)$ is the recombination coefficient 
   (recombination rate from the ion $X_{i+1}$ to $X_i$), $n_{\rm e}$ is the electron number density of the 
   absorbing gas, and $f(X_i)$ is the fraction of ion $X_i$. We select the ions that contribute 
   significantly to the spectral fit for 
   each WA component as the indicators of this component. 
   Some of these parameters can be obtained from {\texttt{SPEX}} code \citep{Mao2016}:
   $\alpha_{\rm{r}}(X_i)$ and $\alpha_{\rm{r}}(X_{i-1})$ come from atomic physics, 
   and $f(X_i)$ and $f(X_{i+1})$ are estimated from the ionization balance. 
   Then we can obtain $nt_{\rm{rec}}$ (see Table \ref{table:table5}).
   The $t_{\rm{rec}}$ of each WA component can be estimated by the variation timescale 
   of the ionization parameters between different observations \citep[e.g.,][]{Ebrero2016b}.
   For WA$_1$, the ionization parameter shows a significant variation between 
   Obs2 and Obs3, so its $t_{\rm{rec}}$ might be lower than the time interval between Obs2 and Obs3 (16 days). 
   For WA$_2$, there is a significant change for the ionization parameter between Obs3 and Obs5, 
   so its $t_{\rm{rec}}$ might be smaller than the time interval between Obs3 and Obs5 (10 days).
   For both WA$_3$ and WA$_4$, Obs3 and Obs4 have a significantly different ionization parameter,
   which indicates that their $t_{\rm{rec}}$ might be smaller than the time interval between Obs3 and Obs4 (6 days).
   However, we cannot make a clear conclusion about the lower limits 
   of the $t_{\rm{rec}}$ for these WA components because of the low number of observations.
   We list the $t_{\rm{rec}}$ of each WA component in Table \ref{table:table5}.
   According to $t_{\rm{rec}}$, $nt_{\rm{rec}}$, and Eq. \ref{equ:xi}
   (we also use the ionizing luminosity of Obs5 here), 
   the radial distance is estimated to be $\lesssim$ 0.16 pc for WA$_1$, 
   $\lesssim$ 0.3 pc for WA$_2$,
   $\lesssim$ 0.6 pc for WA$_3$,
   $\lesssim$ 13 pc for WA$_4$ (see Table \ref{table:table5}).
   These estimates are consistent with the results obtained by 
   Eqs. \ref{equ:rmax} and \ref{equ:rmin} (see Table \ref{table:table5} and Fig. \ref{fig:distance}).

    The radial location of the optical BLR of NGC~3227 is estimated to be 
   around 0.0032 pc from the time 
   lag between the H$\beta$ line and continuum at 5100 $\AA$ \citep{Denney2009}.
   The optical [O III] $\lambda$5007 image indicates that the narrow line region (NLR) of NGC~3227
   can extend to $\sim$ 100 pc \citep{Schmitt1996}, 
   even to $\sim$ 500 pc \citep{Mundell1995}.
   According to \cite{Nenkova2008b}, the inner radius of the torus 
   can be estimated by 
   \begin{equation}
   R_{\rm{in}}=0.4\times \bigg(\frac{L_{\rm{bol}}}{10^{45}\ \rm{erg}\ \rm{s}^{-1}}\bigg)^{0.5} \bigg(\frac{1500\ K}{T_{\rm{d}}}\bigg)^{2.6}\ \rm{pc},
   \label{equ:torusin}
   \end{equation}
   with a dust temperature of $T_{\rm{d}}=1500$ K. 
   If $L_{\rm{bol}}=4.6\times 10^{43}\ {\rm{erg}}\ {\rm{s}}^{-1}$ (see Sect. \ref{sec:SED}),
   $R_{\rm{in}}$ is about 0.09 pc.
   The outer radius can be estimated by $R_{\rm{out}}=Y\times R_{\rm{in}}$.
   Using the clumpy model \citep{Nenkova2002,Nenkova2008a,Nenkova2008b}, 
   \cite{Alonso2011} estimated that the $Y$ of NGC~3227 is about 17,
   so $R_{\rm{out}}$ is around 1.46 pc.
   The distance estimates for 
   the BLR, torus, NLR, and each WA component of NGC~3227 are shown in Fig. \ref{fig:distance}. 

   As Fig. \ref{fig:distance} shows, WA$_1$ lies between the outer region of the BLR and the inner region of torus; 
   WA$_2$ and WA$_3$ might be in the torus.
   WA$_4$ likely resides between the outer torus and the 
   NLR, and it is the one that is farthest away from the AGN compared with other WAs,
   which might explain the lowest ionization parameter of this component 
   (see details in Section \ref{sec:parawa} and Table \ref{table:table3}).
   For the distance estimation of WA$_1$, the first and second methods give inconsistent results,
   which might be due to the following reasons: (1) these two methods use different assumptions,
   which can be expected to have different results;
   (2) we did not consider the possible contribution from the momentum associated with other processes (e.g., magnetic field),
   which might be not negligible for WA$_1$ and 
   might make $C_{\rm{v}}$ (see Eq. \ref{eq:cv}) and $r_{\rm{max}}$ (see Eq. \ref{equ:rmax}) being underestimated. 

   \begin{table}
   \caption{The estimated distances ($r$) of the four WAs (WA$_1$, WA$_2$, WA$_3$, and WA$_4$) of NGC~3227. \label{table:table5}}
   \centering
   \setlength{\tabcolsep}{5pt}
   \begin{tabular}{lcccc}
   \hline\hline\xrowht[()]{10pt}
    Parameter  & WA$_1$ & WA$_2$ & WA$_3$ & WA$_4$ \\
   \specialrule{0.04em}{1pt}{3pt}
\multicolumn{5}{l}{Method 1$^*$: 
Distance $r \ge$ thickness $\Delta r$} \\
   \specialrule{0.00em}{3pt}{3pt}
   $\dot{P}_{\rm{abs}}\ (10^{33}\ \rm{erg}\ \rm{m}^{-1})$$\ ^a$ & 1.68 & 1.44 & 1.54 & 51.3 \xrowht[()]{7pt} \\
   $\dot{P}_{\rm{scat}}\ (10^{32}\ \rm{erg}\ \rm{m}^{-1})$$\ ^b$ & 9.6 & 1.1 & 0.5 & 0.7 \xrowht[()]{7pt} \\
   $C_{\rm{v}}$$\ ^c$ & 0.049 & 0.039 & 0.011 & 0.006 \xrowht[()]{7pt} \\
   $r$ (pc) $^d$  & $\lesssim$ 0.007 & $\lesssim$ 0.24   & $\lesssim$ 0.71      & $\lesssim$ 265 \xrowht[()]{7pt} \\
   \specialrule{0.04em}{1pt}{3pt}
\multicolumn{5}{l}{Method 2$^{\dag}$:  
Outflow velocity $v_{\rm{out}} \ge$ escape velocity $v_{\rm{esc}}$}  \\
   \specialrule{0.00em}{3pt}{3pt}
   $r$ (pc)$^e$ & $\gtrsim$ 0.03 & $\gtrsim$ 0.2   & $\gtrsim$ 0.3      & $\gtrsim$ 4 \xrowht[()]{7pt} \\
   \specialrule{0.04em}{3pt}{3pt}
\multicolumn{5}{l}{Method 3$^{\ddag}$:  
Recombination timescale method}  \\
   \specialrule{0.00em}{3pt}{3pt}
   $\left \langle n_{\rm{e}}t_{\rm{rec}} \right \rangle\ (10^{10}\ {\rm{s}\ {\rm{cm}}^{-3}})$ & 5.6 & 5.4 & 3.9 & 7.2 \xrowht[()]{7pt} \\
   $t_{\rm{rec}}$ (days)$\ ^f$  & $\lesssim$ 16 & $\lesssim$ 10 & $\lesssim$ 6      & $\lesssim$ 6 \xrowht[()]{7pt} \\
   $\left \langle n_{\rm{e}} \right \rangle\ (10^{4}\ {\rm{cm}}^{-3})$$\ ^g$  & $\gtrsim$ 4.0 & $\gtrsim$ 6.3 & $\gtrsim$ 7.5 & $\gtrsim$ 14.0 \xrowht[()]{7pt} \\
   $r$ (pc) $^h$  & $\lesssim$ 0.16 & $\lesssim$ 0.3   & $\lesssim$ 0.6      & $\lesssim$ 13 \xrowht[()]{7pt} \\
   \specialrule{0.04em}{3pt}{1pt}
   \end{tabular}
  \tablefoot{
   \tablefoottext{$*$}
   {The first method to estimate the distances of the WAs, which is based on the 
   assumption that the thickness ($\Delta r$) of the WA cloud does not exceed its distance (r) 
   (see Eqs. \ref{equ:rmax}--\ref{equ:pscat}).}  \\
   \tablefoottext{$a$}{Momentum outflow rate from the radiation being absorbed.}\\
   \tablefoottext{$b$}{Momentum outflow rate from the radiation being scattered.}\\
   \tablefoottext{$c$}{Volume filling factor.}\\
   \tablefoottext{$d$}{The distances of the WAs that are 
   estimated by the first method.}\\
   \tablefoottext{$\dag$}{
   The second method to estimate the distances of the WAs, 
   which is based on the assumption that the outflow velocities of winds ($v_{\rm{out}}$) 
   are larger than or equal to their escape velocities ($v_{\rm{esc}}$) (see Eq. \ref{equ:rmin}).}\\
   \tablefoottext{$e$}{
   The distances of the WAs that are estimated by the second method.}\\
   \tablefoottext{$\ddag$}{
   The third method to estimate the distances of the WAs,
   which is based on the recombination timescale (see Eq. \ref{equ:trec}).}\\
   \tablefoottext{$f$}{Recombination time scale.}\\
   \tablefoottext{$g$}{Electron number density.}\\
   \tablefoottext{$h$}{The distances of the WAs that are estimated by the third method.}\\
  }
   \end{table}

\subsection{Obscurer components}
\label{sec:obscurer}

   According to the softness ratio (see Sect. \ref{sec:spectra}), S6a, S6b, and S1a  
   seem to be in obscured states, which may be caused by 
   the clouds (obscurer components) crossing the line of sight.
   Then we make detailed spectral modellings to check whether the obscurer components are 
   required to explain the significant spectral variation in these slices.
   Firstly, we fixed the parameters of the intrinsic SED and WAs of S1a to 
   those of S1b that is in the unobscured state because we do not expect strong variabilities 
   in the SED shape and WAs parameters on small time scales between S1a and S1b.
   However, the absorption features in the soft X-rays cannot be explained by WAs alone.
   Therefore, we add an obscurer component, which greatly improves the fitting result ($\Delta C \sim 785$).
   Similarly, we firstly fixed the intrinsic SED and WAs parameters of S6b to those of S6c that is in the unobscured state.
   One obscurer component is also required to improve the fitting result with $\Delta C \sim 1218$.
   Adding a second obscurer component cannot improve the fitting result of both S1a and S6b.
   For S6a, we firstly fixed its intrinsic SED and WAs parameters to those of S6c,
   which cannot explain the observational data well.
   We verified that two obscurer components are required to improve the fitting result: $\Delta C \sim 13582$ for adding
   one obscurer component and $\Delta C \sim 31$ for adding a second obscurer component.

\subsubsection{Parameters of the obscurer components}

   The spectral analysis indicates that S6a has two obscurer components: 
   a high-ionization component ($\log\ \xi = 2.81^{+0.15}_{-0.04}$; S6a OC$_{\rm{H}}$) 
   with $C_{\rm{f}}$ of $0.29^{+0.04}_{-0.08}$ and 
   a low-ionization component ($\log\ \xi = 1.02^{+0.25}_{-0.14}$; S6a OC$_{\rm{L}}$) with $C_{\rm{f}}$ of $0.46 \pm 0.03$. 
   S6b and S1a only have one low-ionization obscurer component respectively: 
   $\log\ \xi = 1.89\pm 0.11$ for S6b OC$_{\rm{L}}$ with $C_{\rm{f}}$ of $0.35 \pm 0.03$ and 
   $\log\ \xi = 1.55^{+0.19}_{-0.30}$ for S1a OC$_{\rm{L}}$ with $C_{\rm{f}}$ of $0.21 \pm 0.02$ (see Table \ref{table:table4}). 
   The low-ionization obscurer component (OC$_{\rm{L}}$) has a lower column density ($\sim 10^{22}\ {\rm{cm}}^{-2}$)
   than the high-ionization obscurer component (OC$_{\rm{H}}$; $N_{\rm{H}} \sim 10^{23}\ {\rm{cm}}^{-2}$) (see Table \ref{table:table4}). 
   S6a OC$_{\rm{L}}$ has a larger covering factor than S6a OC$_{\rm{H}}$, considering 1$\sigma$ level uncertainties.
   Our result for S6a is not consistent with that in \cite{Turner2018} 
   which only found one obscurer component with $N_{\rm{H}} \sim 10^{22} \rm{cm}^{-2}$, $\log\ \xi \sim 2$, and $C_{\rm{f}} \sim 60 \%$.
   It might be due to the different SED model and different spectral modelling process.
   The X-ray transmission of each obscurer component in our line of sight to NGC~3227 
   is shown in Fig. \ref{fig:transmission}. OC$_{\rm{L}}$ components can produce absorption features  
   at energies lower than 6 keV, and the OC$_{\rm{H}}$ component mainly absorbs the continuum at energies higher than 0.5 keV.

   \begin{table}
   \caption{The best-fit parameters of the obscurer components for S6a, S6b, and S1a: 
   hydrogen column density ($N_{\rm{H}}$), ionization parameter ($\xi$), covering factor ($C_{\rm{f}}$). \label{table:table4}}
   \centering
   \small
   \setlength{\tabcolsep}{4pt}
   \begin{tabular}{ccccc}
   \hline\hline\xrowht[()]{10pt}
   Comp. & Parameter & S6a & S6b & S1a \\
   \hline\xrowht[()]{10pt}
                  & $N_{\rm{H}}\ (10^{22}\ {\rm{cm}}^{-2})$ & $8.27^{+7.28}_{-1.63}$ & \nodata  & {\nodata} \xrowht[()]{7pt} \\
   OC$_{\rm{H}}$ & $\log\ \xi\ ({\rm{erg\ cm\ s}}^{-1})$ & $2.81^{+0.15}_{-0.04}$ & {\nodata}  & {\nodata} \xrowht[()]{7pt} \\
                  & $C_{\rm{f}}$ & $0.29^{+0.04}_{-0.08}$ & {\nodata}  & {\nodata}  \xrowht[()]{7pt}  \\
    \hline
                  & $N_{\rm{H}}\ (10^{22}\ {\rm{cm}}^{-2})$ & $1.25^{+0.32}_{-0.20}$ & $1.33^{+0.27}_{-0.21}$ & $1.98^{+0.59}_{-0.42}$ \xrowht[()]{7pt} \\
    OC$_{\rm{L}}$ & $\log\ \xi\ ({\rm{erg\ cm\ s}}^{-1})$ & $1.02^{+0.25}_{-0.14}$ & $1.89\pm 0.11$ & $1.55^{+0.19}_{-0.30}$ \xrowht[()]{7pt} \\
                  & $C_{\rm{f}}$ & $0.46\pm 0.03$ & $0.35\pm 0.03$ & $0.21\pm 0.02$ \xrowht[()]{7pt} \\
   \hline
   \end{tabular}
   \tablefoot{ OC$_{\rm{L}}$ is the low-ionization 
obscurer component and OC$_{\rm{H}}$ is the high-ionization obscurer component.}
   \end{table}

\subsubsection{Radial location of the obscurer components}
\label{sec:lococ}

   As we mention in Sect. \ref{sec:occ}, we cannot constrain the outflow velocities of the obscurer components,
   so we cannot estimate the distance of the obscurer components using the methods for the WAs. 
   Here we estimate the distance of the obscurer components based on the crossing time of the obscuring cloud. 
   For simplicity, we assume that an obscuring cloud moves around the 
   central black hole ($M_{\rm{BH}}$) in a circular orbit of radius $R$, so that the keplerian velocity of 
   the obscuring cloud crossing the line of sight is $\varv_{\rm{cloud}}=\sqrt{GM_{\rm{BH}}/R}$. 
   Assuming a spherical geometry for the obscuring cloud, its diameter is $D$ $\sim$ $N_{\rm{H}}/n_{\rm{H}}$ 
   and the size of this cloud crossing the line of sight is approximately equal to $D$, 
   so that $\varv_{\rm{cloud}}$ can also be given by 
   $\varv_{\rm{cloud}}=D/t_{\rm{cross}}=N_{\rm{H}}/(n_{\rm{H}} t_{\rm{cross}})$, 
   where $t_{\rm{cross}}$ is the crossing time of the obscuring cloud. 
   Therefore, we can obtain a relation for $n_{\rm{H}}$:
   \begin{equation}
   n_{\rm{H}}=\frac{N_{\rm{H}}}{t_{\rm{cross}}}\sqrt{\frac{R}{GM_{\rm{BH}}}}
   \label{equ:nh}
   \end{equation}
   Following \cite{Lamer2003}, we combine Eqs. \ref{equ:xi} and \ref{equ:nh} to obtain $R$:
   \begin{equation}
   R=4\times 10^{16} M_{7}^{1/5} \left(\frac{L_{42} t_{\rm{days}}}{N_{22} \xi}\right)^{2/5}\ {\rm{cm}},
   \label{equ:dist}
   \end{equation}
   where $M_{7}=M_{\rm{BH}}/10^{7}\ M_{\odot}$, $L_{42}=L_{\rm{ion}}/10^{42}\ {\rm{erg}}\ {\rm{s}}^{-1}$, 
   $t_{\rm{days}}$ is $t_{\rm{cross}}$ in days, and $N_{22}=N_{\rm{H}}/10^{22}\ {\rm{cm}}^{-2}$. 

   On the one hand, Obs5 is in an unobscured state, which indicates that OC$_{\rm{H}}$ and OC$_{\rm{L}}$ 
   of S6a do not start to cross the line of sight during the observational time of Obs5. One the other hand,
   OC$_{\rm{H}}$ of S6a disappears in the observational time of S6b. Therefore, $t_{\rm{cross}}$ of OC$_{\rm{H}}$
   should be larger than the exposure time of S6a (39 ks) and smaller than the time interval between Obs5 and Obs6 (309 ks).
   If we assume that S6a and S6b have different OC$_{\rm{L}}$ components, 
   $t_{\rm{cross}}$ of S6a OC$_{\rm{L}}$ should be also between 39 and 309 ks (similar to the case of S6a OC$_{\rm{H}}$), 
   and $t_{\rm{cross}}$ of S6b OC$_{\rm{L}}$ should be comparable to the exposure time of S6b (20 ks).
   Similarly, $t_{\rm{cross}}$ of S1a OC$_{\rm{L}}$ should be larger than the exposure time of S1a (20 ks), 
   and smaller than the time interval (372 days) between Obs1 and 
   an unobscured observation taken in November 2005
   \citep[see Fig. A12 of][]{Markowitz2014}.
   The estimated crossing time of each obscurer component is summarized in Table \ref{table:table6}.
   Then we can use Eq. \ref{equ:dist} to constrain the location of each obscurer component, 
   which is shown in Table \ref{table:table6} and Fig. \ref{fig:distance}.
   These obscurer components are estimated to be located within the BLR,
   which is consistent with the results in previous works \citep{Lamer2003,Beuchert2015,Turner2018}. 

   The obscurers of NGC~3227 are closer to the SMBH than the WAs 
   and also have a significantly larger hydrogen or electron number 
   density than the WAs (see Figure \ref{fig:distance}).
   Besides that, the obscurers usually appear on a short timescale 
   while WAs can exist for a long time,
   and the appearance of the obscurers mainly affects the ionization state of the WAs
   on the short timescale.
   These results indicate that the obscurers and WA outflows might have different origin.
   For example, obscurers might be triggered by the collapse of 
   inner BLR clouds \citep{Kriss2019a,Kriss2019b,Devereux2021}, 
   while WA outflows might be formed by the outflowing of the clouds between outer 
   BLR and NLR under the drive of the radiation pressure \citep[e.g.,][]{Proga2004}, 
   magnetic forces \citep[e.g.,][]{Blandford1982, Konigl1994, Fukumura2010}, 
   or thermal pressure \citep[e.g.,][]{Begelman1983, Krolik1995, Mizumoto2019}.
   However, we cannot get a solid conclusion in this work, 
   which might require more high-quality data to investigate.

   \begin{table}
   \caption{The crossing time ($t_{\rm{cross}}$) and distances ($R$) of the obscurer components for S6a, S6b, and S1a. \label{table:table6}}
   \centering
   \setlength{\tabcolsep}{12pt}
   \begin{tabular}{ccccc}
   \hline\hline\xrowht[()]{10pt}
    Comp.                     & $t_{\rm{cross}}$ & $R$ (pc) \\
   \hline\xrowht[()]{10pt}
   S6a OC$_{\rm{H}}$ & 39--305 ks & 0.001--0.002  \xrowht[()]{7pt} \\
   S6a OC$_{\rm{L}}$ & 39--305 ks & 0.011--0.02  \xrowht[()]{7pt} \\
   S6b OC$_{\rm{L}}$ & $\sim$ 20 ks & $\sim$ 0.004   \xrowht[()]{7pt} \\
   S1a OC$_{\rm{L}}$ & 20 ks--372 days & 0.004--0.07 \xrowht[()]{7pt} \\
   \hline
   \end{tabular}
   \end{table}

\section{SUMMARY AND CONCLUSIONS}
\label{sec:summary}

  The relation between WA outflows of AGN 
  and nuclear obscuration activities is still unclear.
  NGC~3227 is a suitable target to study the 
  properties of both WAs and obscurers, 
  which might help us understand their correlation.
   To investigate the WA components of NGC~3227 in detail, we use a broadband SED model (Paper I)
   and the photoionization model in \texttt{SPEX} software to fit the unobscured spectra of the archival and previously 
   published {\it XMM-Newton} and {\it NuSTAR} observations taken in 2006 and 2016.
   Based on the broadband SED and WAs parameters, 
   we also study the X-ray obscuration events in the archival observations.

   We detect four ionization phases for the WAs in NGC~3227 using the unobscured observations: 
   $\log\ \xi\ ({\rm{erg}}\ {\rm{cm}}\ {\rm{s}}^{-1}) \sim -1.0,\ 2.0,\ 2.5,\ 3.0$. The highest-ionization 
   WA component has a much higher hydrogen column density ($N_{\rm{H}} \sim 10^{22} \ {\rm{cm}}^{-2}$) 
   than the other three WA components ($N_{\rm{H}} \sim 10^{21}\ {\rm{cm}}^{-2}$). The outflow 
   velocities of these WA components range from $\sim -100$ to $\sim -1300$ ${\rm{km}}\ {\rm{s}}^{-1}$, and show a positive 
   correlation with the ionization parameter. 
   Our estimates of the radial location of these WA components indicate 
   that the WAs of NGC~3227 might reside over radii ranging from the BLR to the torus, even to the NLR.

   We find an obscuration event in 2006, which was missed by previous studies. 
   One obscurer component is required for this 2006 obscuration event. 
   For the previously published obscuration event in 2016, we detect two obscurer components. 
   A high-ionization obscurer component ($\log\ \xi = 2.81_{-0.04}^{+0.15}$; 
   covering factor $C_{\rm{f}}=0.29^{+0.04}_{-0.08}$) only appears in the 2016
   observation, which has a column density around $10^{23}\ \rm{cm}^{-2}$, 
   while both the 2006 and 2016 observations have a low-ionization obscurer component 
   ($\log\ \xi \sim$ 1.0--1.9; $C_{\rm{f}} \sim 0.2-0.5$), 
   which has a lower column density ($\sim 10^{22}\ \rm{cm}^{-2}$) than the high-ionization obscurer component.
   Assuming that the variations of flux is caused by the 
   transverse motion of obscurers across the line of sight, we estimate the locations of obscurers to be within the BLR.

   The obscurers of NGC~3227 are closer to the SMBH than the WAs 
   and have a significantly larger hydrogen or electron number 
   density than the WAs.
   In addition, the obscurers usually appear on a short timescale while the WAs can exist for a long time.
   These proofs indicate that the obscurers and WAs of NGC~3227 might have different origins.

\begin{acknowledgements}
      We thank the referee for helpful comments that improved this paper. 
      This research has made use of the NASA/IPAC 
      Extragalactic Database (NED), which is funded by the National Aeronautics 
      and Space Administration and operated by the California Institute of Technology.
      YJW gratefully acknowledges the financial support from the China Scholarship Council.
      YJW and YQX acknowledge support from NSFC-12025303, 11890693, 11421303, 
      the CAS Frontier Science Key Research Program (QYZDJ-SSW-SLH006), 
      the K.C. Wong Education Foundation, and the science research grants from the 
      China Manned Space Project with NO. CMS-CSST-2021-A06.
      SRON is supported financially by NWO, the Netherlands Organization for Scientific Research.
      JM acknowledges the support from STFC (UK) through the 
      University of Strathclyde UK APAP network grant ST/R000743/1.
      GP acknowledges funding from the European Research Council (ERC) under the European 
      Union's Horizon 2020 research and innovation programme (grant agreement No 865637). 
      E.B. is funded by a Center of Excellence of THE ISRAEL SCIENCE FOUNDATION (grant No. 2752/19).
      SB acknowledges financial support from ASI under grants ASI-INAF I/037/12/0 and n. 2017-14-H.O., 
      and from PRIN MIUR project "Black Hole winds and the Baryon Life Cycle of Galaxies: 
      the stone-guest at the galaxy evolution supper", contract no. 2017PH3WAT.
      BDM acknowledges support via Ram{\'{o}}n y Cajal Fellowship RYC2018-025950-I.
      SGW acknowledges the support of a PhD studentship awarded by 
      the UK Science \& Technology Facilities Council (STFC).
      DJW acknowledges support from STFC in the form of an Ernest Rutherford fellowship.
      P.O.P acknowledges financial support from the CNES french agency 
      and the PNHE high energy national program of CNRS.

\end{acknowledgements}

\bibliographystyle{aa}
\bibliography{NGC3227_WA_ms.bbl}

\end{CJK*}
\end{document}